\documentclass[journal=jpcafh,manuscript=article]{achemso}

\usepackage[version=3]{mhchem} 
\usepackage[utf8]{inputenc}
\usepackage{chemfig}
\usepackage[T1]{fontenc} 
\usepackage{graphicx}
\usepackage{dcolumn}
\usepackage{siunitx}
\DeclareSIUnit\angstrom{\text{Å}}
\usepackage{amsmath}
\usepackage{cmbright}
\usepackage{multirow}
\usepackage{xcolor}
\usepackage[ruled]{algorithm2e}
\usepackage{hyperref}
\hypersetup{
    colorlinks=true,
    linkcolor=cyan,
    filecolor=magenta,      
    urlcolor=cyan,
    citecolor=cyan,
    }
    
\DeclareMathVersion{sans}
\SetSymbolFont{operators}{sans}{OT1}{cmbr}{m}{n}
\SetSymbolFont{letters}{sans}{OML}{cmbr}{m}{it}
\SetSymbolFont{symbols}{sans}{OMS}{cmbr}{m}{n}
\SetMathAlphabet{\mathit}{sans}{OT1}{cmbr}{m}{sl}
\SetMathAlphabet{\mathbf}{sans}{OT1}{cmbr}{bx}{n}
\SetMathAlphabet{\mathtt}{sans}{OT1}{cmtl}{m}{n}
\SetSymbolFont{largesymbols}{sans}{OMX}{cmbr}{m}{n}

\mathversion{sans}

\SectionNumbersOn

\usepackage{textgreek}

\author{Jasmine Bone}
\affiliation[UB]{Centre for Computational Chemistry, School of Chemistry, University of Bristol, Bristol BS8 1TS, United Kingdom}
\author{Javier Carmona-Garc\'{i}a}
\affiliation[UB]{Centre for Computational Chemistry, School of Chemistry, University of Bristol, Bristol BS8 1TS, United Kingdom}
\author{Daniel Hollas}
\email{daniel.hollas@bristol.ac.uk}
\affiliation[UB]
{Centre for Computational Chemistry, School of Chemistry, University of Bristol, Bristol BS8 1TS, United Kingdom}
\author{Basile F. E. Curchod}
\email{basile.curchod@bristol.ac.uk}
\affiliation[UB]
{Centre for Computational Chemistry, School of Chemistry, University of Bristol, Bristol BS8 1TS, United Kingdom}

\title{Benchmarking electronic-structure methods for the description of dark transitions in carbonyls at and beyond the Franck-Condon point}

\begin{document}

\begin{tocentry}
\includegraphics[width=\textwidth]{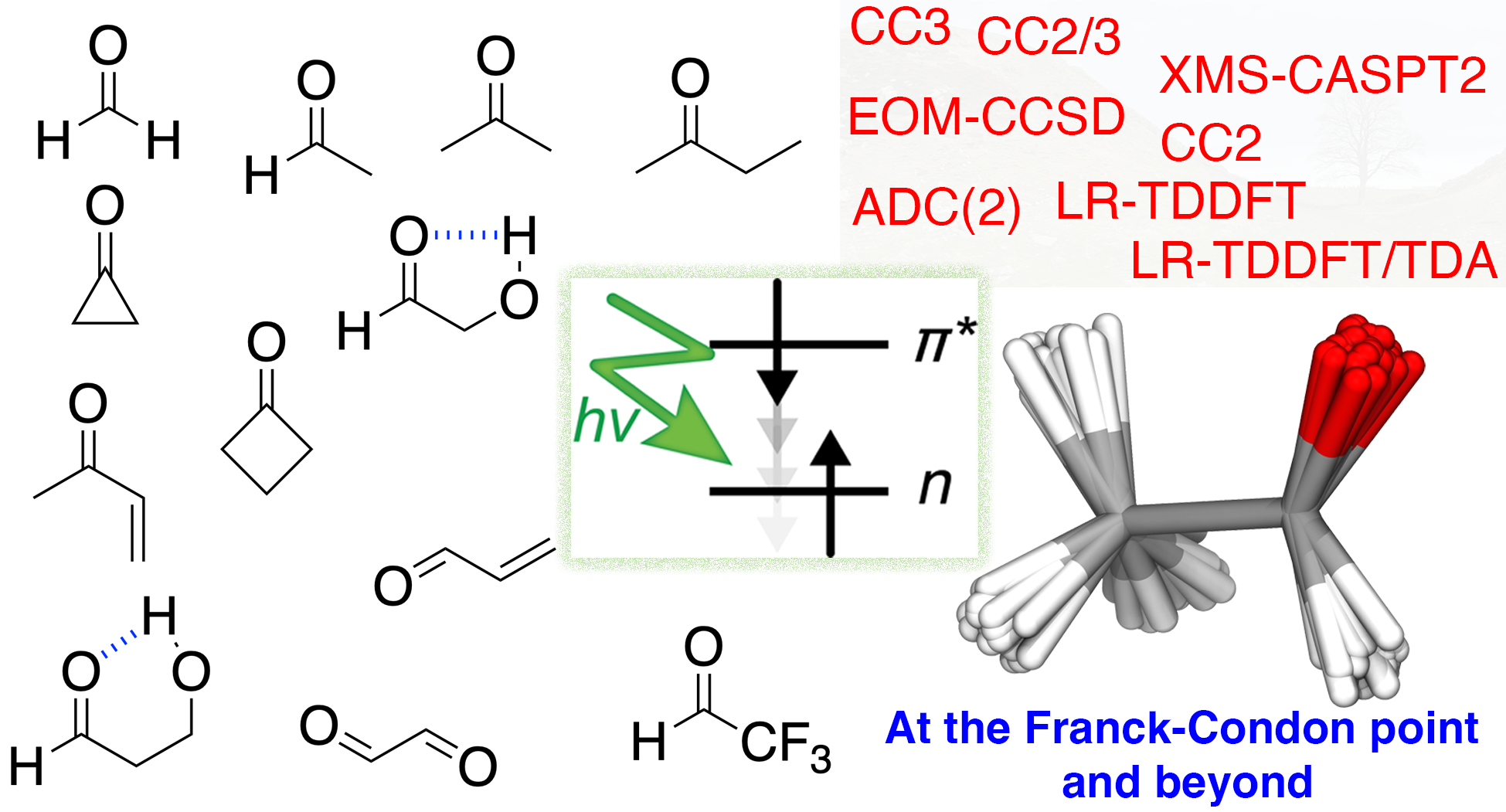}

\end{tocentry}

\begin{abstract}
Herein, we propose a comprehensive benchmark of electronic-structure methods to describe dark transitions, that is, transitions to excited electronic states characterized by a near-zero oscillator strength. This type of electronic state is particularly important for the photochemistry of molecules containing carbonyl groups, such as atmospheric volatile organic compounds (VOCs). The oscillator strength characterizing a dark transition can change dramatically by a slight alteration of the molecular geometry around its ground-state equilibrium, the so-called non-Condon effects. Hence, testing the performance of electronic-structure methods for dark transitions requires considering molecules at their Franck-Condon point (i.e., equilibrium geometry), but also beyond the Franck-Condon point.
Our benchmark focuses on various electronic-structure methods -- LR-TDDFT(/TDA), ADC(2), CC2, EOM-CCSD, CC2/3, XMS-CASPT2 -- with CC3/aug-cc-pVTZ serving as a theoretical best estimate. These techniques are tested against a set of 16 carbonyl-containing VOCs at their equilibrium geometry. We then assess the performance of these methods to describe the dark transition of acetaldehyde beyond its Franck-Condon point by (i) distorting the molecule towards its S$_1$ minimum energy structure and (ii) sampling an approximate ground-state quantum distribution for the molecule and calculating photoabsorption cross-sections within the nuclear ensemble approach. Based on the calculated cross-sections, we calculate the photolysis half-life as depicted by the different electronic-structure methods -- highlighting the impact of the different electronic-structure methods on predicted experimental photolysis observables.  
\end{abstract}

\section{Introduction}

Determining effective methods for the reliable prediction of molecular properties for excited electronic states provides a major challenge for theoretical chemistry, lending itself towards the fact that ground-state methods are usually more accurate than their excited-state analogues.\cite{https://doi.org/10.1002/wcms.1517} Over the past two decades, numerous systematic efforts have emerged to benchmark electronic-structure methods for excited electronic states, focusing on calculating and comparing excited-state properties such as vertical excitation energies and oscillator strengths using curated datasets. The most notable of these efforts are those forming the Thiel’s set,\cite{10.1063/1.2889385, 10.1063/1.2973541, Silva-Junior10022010} QUEST databases,\cite{https://doi.org/10.1002/wcms.1517, loos2018mountaineering,doi:10.1021/acs.jctc.9b01216, doi:10.1021/acs.jctc.8b01205, doi:10.1021/acs.jctc.0c00227} and Gordon’s set.\cite{10.1063/1.3689445, 10.1063/5.0018354, doi:10.1021/acs.jctc.9b00013, doi:10.1021/acs.jctc.7b00386}  These sets contain hundreds of vertical excitation energies calculated across a wide variety of molecular systems obtained with methods such as linear-response time-dependent density-functional theory (LR-TDDFT),\cite{Runge84,casida95,petersilka96} algebraic diagrammatic construction (ADC(n)),\cite{dreuw2015algebraic} equation-of-motion coupled-cluster (EOM-CC),\cite{eom} approximate coupled-cluster methods (CC$_x$),\cite{sneskov2012excited} and configuration interaction (CI).\cite{EVANGELISTI198391} They also include some multireference methods like complete active space second order perturbation theory (CASPT2)\cite{BATTAGLIA2023135} and N-electron valence state second order perturbation theory (NEVPT2).\cite{nevpt2} Throughout these studies, CC3 was consistently identified as the most reliable method applicable to sizeable molecules with transitions exhibiting a single-excitation character.\cite{Loos_2020} 

Traditional benchmark studies, like those discussed above, have laid essential groundwork for the validation of excited-state electronic-structure methods. However, most benchmark efforts have focused on bright, symmetry-allowed excitations at the Franck-Condon (FC) point (i.e., at the optimized ground-state equilibrium geometry), leaving a critical gap in the reliable prediction of transitions with near-zero oscillator strengths, the so-called dark transitions. Dark, symmetry-forbidden transitions, with $n\pi^\ast$ excitation being a prototypical example, are characterized by small oscillator strengths (typically f < 0.01) and, as such, have limited immediate spectroscopic signatures. Earlier benchmarks have focused on excitation energies for dark ($n\pi^\ast$) transitions (e.g., Refs.~\citenum{loos2018mountaineering,doi:10.1021/acs.jctc.9b01216,loos2024mountaineering}), meaning that a benchmark for their oscillator strength is not generally available. Perhaps more importantly, the oscillator strengths for $n\pi^\ast$ transitions are known to be highly sensitive to nuclear geometry, particularly along low-frequency vibrational modes that distort the system away from its high symmetry,\cite{turro2009principles} leading to increases in oscillator strengths in nuclear configurations away from the FC point. These effects, which break the Condon approximation, require a benchmark of electronic-structure methods beyond the FC point to ensure the consistency of the method in describing excitation energies and, more importantly, oscillator strengths -- pivotal for an accurate depiction of absorption spectra or photoabsorption cross-sections. 

A typical example where dark transitions play a major role is atmospheric chemistry. 
Carbonyl-containing volatile organic compounds (VOCs), such as aldehydes and ketones, are highly abundant in the atmosphere, emitted both directly from anthropogenic and biogenic sources and formed as secondary products.\cite{LIU2022106184}. These molecules can absorb sunlight in the actinic region (280-400 nm) via $n\pi^\ast$ transitions, inducing complex photochemical reactivity that leads to radical formation and contributes to processes such as ozone production and secondary organic aerosol (SOA) formation. 
The photolysis rate coefficient, $j$, of a given VOC is given by  
\begin{equation}
\textit{j} = \int_{\lambda_{min}}^{\lambda_{max}} \sigma(\lambda) \phi(\lambda) F(\lambda) d\lambda \, ,
\label{eq:photolysis}
\end{equation}
where F($\lambda$) is the flux of the light source, $\phi({\lambda})$ is the (wavelength-dependent) quantum yield for the photolysis process of interest, and $\sigma({\lambda})$ is the photoabsorption cross-section of the VOC. Modelling chemical transformations in the troposphere requires the determination of photolysis rate coefficients for (transient) VOCs, which can be challenging to determine experimentally due to the instability of these compounds.\cite{acp-24-13317-2024,Curchodatmophotochem2024} Recent efforts have attempted to use computational photochemistry to predict photoabsorption cross-sections (and quantum yields) for atmospheric molecules (e.g., Refs.~\citenum{carmona-garcia2021hoso,rocasanjuan2020mercury,Prljphotolysis2020,hollas2024atmospec}). Special attention was given to including nuclear quantum effects to adequately capture the shape of the photoabsorption cross-section, $\sigma(\lambda)$, by lifting the Condon approximation, that is, considering the variation in transition dipole moment induced by the distortion of a molecule away from its equilibrium geometry (FC point).\cite{C8CP00199E,prlj2021calculating,prlj2023deciphering}
Yet, a quantitative description of a photoabsorption cross-section ultimately depends on the capabilities of the underlying electronic-structure method to provide accurate excitation energies and oscillator strengths for these dark transitions, at and beyond the FC point. Such knowledge is currently lacking in the literature. 

In this work, we propose a focused benchmarking study on dark transitions in representative carbonyl-containing molecules (depicted in Figure~\ref{fig:struct}), using a suite of single- and multireference electronic structure methods, including LR-TDDFT(/TDA), ADC(2), EOM-CCSD, CC2, XMS-CASPT2, and the composite method, CC2/3. CC3 is used as a reference. The capability of these methods to provide accurate vertical excitation energies and oscillator strengths is first tested in the FC region for the selected carbonyl-containing molecules. A selected compound, acetaldehyde, is then used to challenge the accuracy of each electronic-structure method beyond the FC point by (i) determining excitation energies and oscillator strengths along a path connecting the optimized ground-state (S$_0$) geometry to the optimized geometry of the first excited electronic state, S$_1(n\pi^\ast)$, (ii) calculating excitation energies and oscillator strengths for a set of 50 geometries sampled from an approximate ground-state distribution, and (iii) comparing the predicted theoretical photoabsorption cross-sections ($\sigma(\lambda)$) and photolysis half-lifes ($\frac{ln(2)}{j}$, with $j$ from Eq.~\eqref{eq:photolysis}). Our results highlight the importance of considering geometries beyond the FC point for benchmarks of electronic-structure methods for excited electronic states and provide guidance for the selection of an adequate electronic-structure method for the study of molecules exhibiting dark transitions.  

\section{Methodology}
\label{sec:method}
\subsection{Molecules and geometries considered -- Franck-Condon point and beyond}

 \begin{figure}[H]
     \centering
     \includegraphics[width=0.7\textwidth]{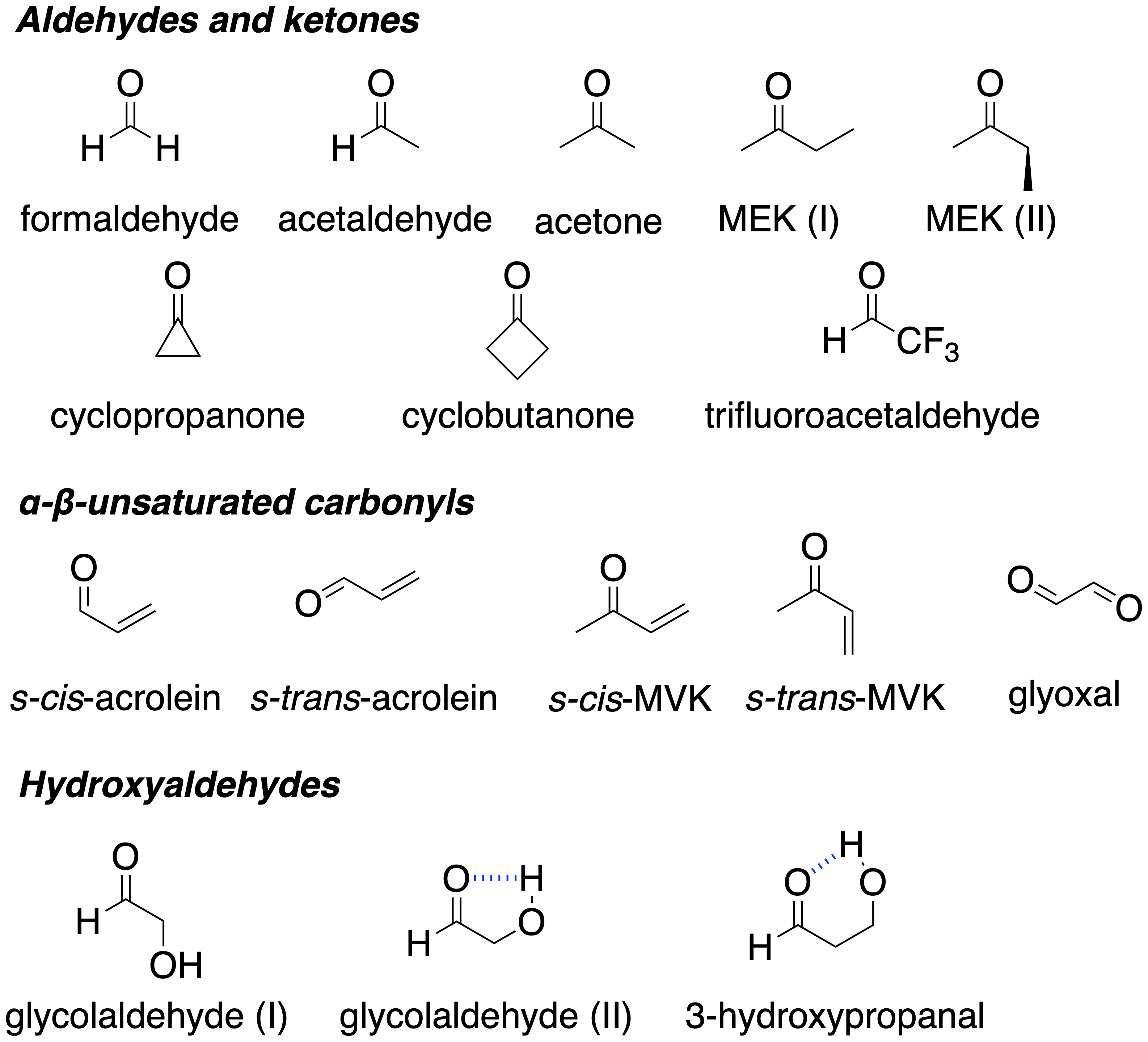}
     \caption{Carbonyl-containing molecules used for this benchmark study. Labels '(I)' or '(II)' identify different conformers of the same molecule. MEK = methyl ethyl ketone; MVK = methyl vinyl ketone.}
     \label{fig:struct}
 \end{figure}

The selected benchmark set of 16 medium-sized carbonyl-containing molecules (Figure \ref{fig:struct}), comprising aldehydes, ketones, $\alpha$-$\beta$-unsaturated carbonyls, and hydroxyaldehydes, is intended to cover many types of carbonyl motifs of importance in VOCs found in the troposphere. Formaldehyde, cyclopropanone, and acetone are symmetric and planar, resulting in a precisely zero oscillator strength for the transition to the S$_1(n\pi^\ast)$ electronic state at the ground-state equilibrium geometry. Other carbonyl compounds provide an opportunity to test the role played by different conformations or hydrogen bonds on the excitation energy and oscillator strength of the S$_0\rightarrow$S$_1(n\pi^\ast)$ transition. We note that the lowest $n\pi^\ast$ transition energy exhibited by some of these molecules was used in earlier benchmarking studies.\cite{loos2018mountaineering,doi:10.1021/acs.jctc.9b01216,loos2024mountaineering} 

Unless otherwise stated, the same computational protocol was employed for all molecules. The ground-state geometries of these molecules were optimized using M{\o}ller-Plesset second-order perturbation theory (MP2)\cite{PhysRev.46.618} with the cc-pVTZ basis set within the ORCA package (v5.0.4).\cite{10.1002/wcms.1606}. A frequency calculation at the same level of theory was performed to confirm that the extremum located is a minimum on the ground-state potential energy surface. The equilibrium geometries obtained at the MP2/cc-pVTZ level of theory, the \textit{Franck-Condon point} for each molecule, were used as the ground-state reference geometry for all single-point excited-state calculations discussed in Section~\ref{sec:FCP}.

The geometry of acetaldehyde was further optimized in the S$_1$($n\pi^*$) state with CC2/cc-pVTZ, and the nature of the S$_1$ minimum located was confirmed by a frequency calculation. The choice of CC2 for the geometry optimization (in place of ADC(2)) was motivated by the reported poor performance of ADC(2) in describing the potential energy surface for electronic states with $n\pi^\ast$ character.\cite{marsili2021caveat} A linear interpolation in internal coordinates (LIIC) was conducted between the optimized S$_0$ (MP2/cc-pVTZ) and optimized S$_1$ (CC2/cc-pVTZ) geometries of acetaldehyde, yielding a set of ten geometries connecting these two critical points. The LIIC pathway obtained represents a set of twelve geometries describing a path dictated by an interpolation (not a minimum-energy path) from the FC point to the S$_1$ minimum, and is used to benchmark electronic-structure methods beyond the FC region (Section~\ref{sec:beyondFCP}). 

To assess the importance of small geometrical changes around the FC point, a series of molecular geometries were sampled from a ground-state nuclear distribution of acetaldehyde at 0\,K. The ground-state nuclear distribution was approximated by a harmonic Wigner distribution, built from the S$_0$ optimized geometry of acetaldehyde and corresponding normal modes. An optimal sampling approach\cite{srsen2021optimalrepNEA} implemented in the PyNEAppLES package\cite{pyneapples} was used to reduce the initial set of 4000 sampled geometries to a smaller number of geometries, here 50, playing the most critical role in reproducing the photoabsorption cross-section of acetaldehyde. This smaller number of geometries is amenable to CC3 calculations, used as a reference as further discussed below using the nuclear ensemble approach (NEA, additional information about this method is provided below). Photoabsorption cross-sections and photolysis rate coefficients were calculated with AtmoSpec.\cite{hollas2024atmospec} The calculation of the photolysis rate coefficient for a given electronic-structure method used its predicted photoabsorption cross-section, a (wavelength-independent) quantum yield of 1.0, and the standardized medium actinic flux (solar zenith angle = 60$^\circ$, overhead ozone column = 350 DU) for a ground elevation of 0 km above sea level.

The optimized geometries for all the molecules considered in this work, as well as the LIIC pathway and Wigner-sampled geometries for acetaldehyde can be found as part of the Supporting Information (SI). We also provide template input files and all the electronic energies and oscillator strengths calculated.

\subsection{Basis set}

A broad range of standard basis sets was tested, namely Dunning’s cc-pVXZ and aug-cc-pVXZ ( X = D, T, and Q) basis sets\cite{dunning89, kendall92} and Ahlrich’s def2-XVP and def2-XVPD (X = S, TZ, and QZ) basis sets.\cite{B508541A, B515623H} Augmented basis functions were found to be important for the description of oscillator strengths for the dark state considered (see SI), spotlighting aug-cc-pVDZ and aug-cc-pVTZ as promising candidates for our benchmark. Tests along the LIIC pathway revealed that aug-cc-pVTZ was needed to ensure an adequate description of excitation energies and oscillator strengths in and beyond the FC point (SI). Based on these findings, the aug-cc-pVTZ basis set was used for all calculations presented in this work. Additional details on the basis-set benchmarking are provided in the SI.

\subsection{Electronic-structure methods}

Our benchmark focuses solely on the lowest excited electronic state (S$_1$) with an $n\pi^*$ character for all molecules considered. 

Linear-response time-dependent density functional theory (LR-TDDFT) within the adiabatic approximation\cite{tddftcarsten} was tested in combination with four broadly-used functionals: B3LYP\cite{Becke1988a, b3}, PBE0 \cite{AB99}, CAM-B3LYP,\cite{Yanai04} and $\omega$B97X-D4.\cite{B810189B, 10.1063/1.2834918} LR-TDDFT calculations can be performed in the full linear-response regime or using the Tamm-Dancoff approximation (TDA). TDA is known to reduce the computational demand of LR-TDDFT, yielding excitation energies close to corresponding full LR-TDDFT calculations.\cite{hirata99}. Yet, this reduction in computational cost sometimes comes with a deterioration of oscillator strengths,\cite{chantzis2013tamm} rationalized by the fact that TDA breaks the Thomas-Reiche-Kuhn sum rule.\cite{furche01,hutter03} Given the utmost importance of adequately describing oscillator strengths when calculating photoabsorption cross-sections for carbonyl-containing atmospheric molecules, we tested LR-TDDFT both with and without TDA in the following. For all LR-TDDFT(/TDA) calculations, the first three lowest excited states were considered, using oscillator strengths within the length gauge (see Ref.~\citenum{doi:10.1021/acs.jctc.0c01228} for a discussion and comparison of gauges for oscillator strengths). All LR-TDDFT(/TDA) calculations were conducted within the ORCA (v5.0.4) package. \cite{10.1002/wcms.1606}

The performance of a range of single-reference (wavefunction-based) methods -- ADC(2) (second-order algebraic diagrammatic construction), CC2 (coupled-cluster singles and approximate doubles),\cite{christiansen95,sneskov2012excited} EOM-CCSD (equation-of-motion coupled-cluster singles and doubles),\cite{eom} CC3 (coupled-cluster singles, doubles and perturbative triples)\cite{10.1063/1.473322, sneskov2012excited} -- was tested to evaluate their accuracy in predicting vertical excitation energies and oscillator strengths for dark, low-lying excited states. CC3 was selected here as our reference due to its ability to offer highly accurate vertical transition energies and oscillator strengths for transitions with a dominant single-excitation character, \cite{doi:10.1021/acs.jctc.9b01216, doi:10.1021/acs.jctc.7b01224} despite its computational cost.\cite{doi:10.1021/acs.jctc.0c00686} All CC based calculations in this work were performed using the eT (v1.9) program package,\cite{10.1063/5.0004713} and all ADC(2) calculations were performed using the ORCA software package (v5.0.4) \cite{10.1002/wcms.1606}.

Multireference calculations were performed with the extended multi-state complete-active-space second-order perturbation theory (XMS-CASPT2) method\cite{BATTAGLIA2023135, 10.1063/1.3633329, 10.1021/j100377a012}. The active space generally included all $\pi$ and $\pi^*$ orbitals, as well as any \textit{n} orbitals present. The $\sigma$/$\sigma^*$ orbitals of the carbonyl moiety were excluded as they proved to have no impact on the low-lying electronic state of interest. The smallest reasonable choice of active space consists of 4 electrons in 3 orbitals (\textit{n}, $\pi$ and $\pi^*$). Extensions of this active space were tested for each individual molecule. The (4,3) active space was used for acetaldehyde, 3-hydroxypropanal, acetone, trifluoroacetaldehyde, formaldehyde, MEK, cyclobutanone, cyclopropanone, and glycolaldehyde.
The (6,5) active space, used for acrolein and MVK, included contributions from the $\pi$/$\pi^*$ orbitals of the alkene bond as well as the $n$, $\pi$, and $\pi^*$ orbitals from the carbonyl. The (8,6) active space -- used solely for glyoxal -- included the $n$, $\pi$, and $\pi^*$ orbitals, both in phase and out of phase, to describe the excitation character in both carbonyls present within the compound. The state averaging process considered two electronic states (S$_0$ and S$_1$), except for glyoxal where a third electronic state (S$_2$) was considered due to the presence of a second $n\pi^\ast$ state close in energy to the S$_1(n\pi^\ast)$ state.\cite{glyoxalelectronicstructure} XMS-CASPT2 was employed within the single-state single reference (SS-SR) contraction scheme, with frozen core and density fitting approximations (using the cc-pVTZ-jkfit basis set from the BAGEL library\cite{bagel}) also applied. The occurrence of intruder states was checked by inspection of the weight of the reference function in the perturbation treatment. An imaginary shift of 0.1 a.u. was applied to all XMS-CASPT2 calculations. All XMS-CASPT2 calculations reported in this work were conducted using BAGEL 1.2.2 package.\cite{bagel} A few tests were performed with OpenMolcas (v24.10)\cite{openm} as well for acetaldehyde (using the same imaginary shift, active space, and basis set) and presented in the following.

As stated above, the reference CC3 can rapidly become too computationally demanding for molecules larger than those in our benchmark set. We tested the accuracy of a composite method, coined CC2/3, which showed great success in reproducing the photoabsorption cross-section of Criegee intermediates.\cite{C8CP00199E} In short, we used CC2/3 to correct the results of a CC2/aug-cc-pVTZ calculation for the role of the perturbative triples by using a CC3 calculation with a smaller basis set, namely def2-SVPD. CC2/3 excitation energies are obtained as 
\begin{equation}
E_{\text{CC2/3}} = E^{\text{aug-cc-pVTZ}}_{\text{CC2}} + (E^{\text{def2-SVPD}}_{\text{CC3}} - E^{\text{def2-SVPD}}_{\text{CC2}}) \, 
\end{equation}
and CC2/3 oscillator strengths ($f$) are determined from
\begin{equation}
f_{\text{CC2/3}} = f^{\text{aug-cc-pVTZ}}_{\text{CC2}}\times \frac{f^{\text{def2-SVPD}}_{\text{CC3}}}{f^{\text{def2-SVPD}}_{\text{CC2}}} \, .
\end{equation}
A similar approach was employed in Ref.~\citenum{loos2022mountaineering} to estimate CC4/aug-cc-pVTZ energies from CCSDT and CC4 with a smaller basis set. 
A discussion on the impact of the basis set size on the CC2/3 results can be found in the SI. 

\subsection{Statistical measures}
Various statistical measures are used in subsequent Sections of this paper to capture the behavior of each electronic-structure method tested: the mean signed error (MSE), 
\begin{equation}
\text{MSE} = \frac{1}{N}\sum\limits^{N}_{i=1}\Delta x_i \, ,
\end{equation}
the mean absolute error (MAE), 
\begin{equation}
\text{MAE} = \frac{1}{N}\sum\limits^{N}_{i=1} |\Delta x_i|  \, ,
\end{equation}
the standard deviation of the errors (SDE), 
\begin{equation}
\text{SDE} = \sqrt{\frac{1}{N-1}\sum\limits^{N}_{i=1}(\Delta x_i - \text{MSE})^2} \, ,
\end{equation}
and the maximum error (MAX) 
\begin{equation}
\text{MAX} = \max_{i} |\Delta x_i| \, .
\end{equation}
In the previous definitions, $\Delta x_i = x_i^\text{calc} - x_i^\text{ref}$ where $x_i^\text{calc}$ refers to the value of interest (excitation energy or oscillator strength) obtained with a given method for molecule $i$ and $x_i^\text{ref}$ is the reference value for this molecule $i$ obtained at the CC3/aug-cc-pVTZ level of theory. $N$ is the total number of molecules considered. 

MSE assesses the accuracy of excitation energy and oscillator strength estimation by a given method, giving an idea about method bias. A positive value for MSE indicates that a method tends to overestimate values on average, and a negative MSE denotes that the method tends to underestimate values on average. MAE calculates the average magnitude of errors between predicted and observed values. Small values for MAE indicate better method accuracy. SDE is used to measure the spread of variability of errors around the mean -- it captures the consistency of a method. High values of SDE indicate that a method is less reliable with increased fluctuations in error, case by case. It is worth noting that SDE can be distorted by extreme outliers. MAX provides the largest absolute difference between predicted and actual values and acts as a measure for the 'worst case error'. 

\section{Results and Discussion}

Section~\ref{sec:FCP} describes a benchmark of excitation energies and oscillator strengths as depicted by different electronic-structure methods for the lowest dark transition ($n\pi^\ast$) of a set of 16 carbonyl-containing molecules (depicted in Figure~\ref{fig:struct}) at their FC point. Section~\ref{sec:beyondFCP} focuses on acetaldehyde and tests the performance of electronic-structure methods in describing non-Condon effects for its dark transition by distorting the molecule towards its S$_1(n\pi^\ast)$ minimum. Section~\ref{sec:FAP} finally assesses the performance of each electronic-structure method in describing the dark transition of acetaldehyde for a range of molecular geometries representative of a ground-state distribution, used to predict a photoabsorption cross-section.
 
\subsection{Benchmark at the Franck-Condon point}
\label{sec:FCP}
\begin{figure}[h!]
    \centering
    \includegraphics[width=1.0\textwidth]{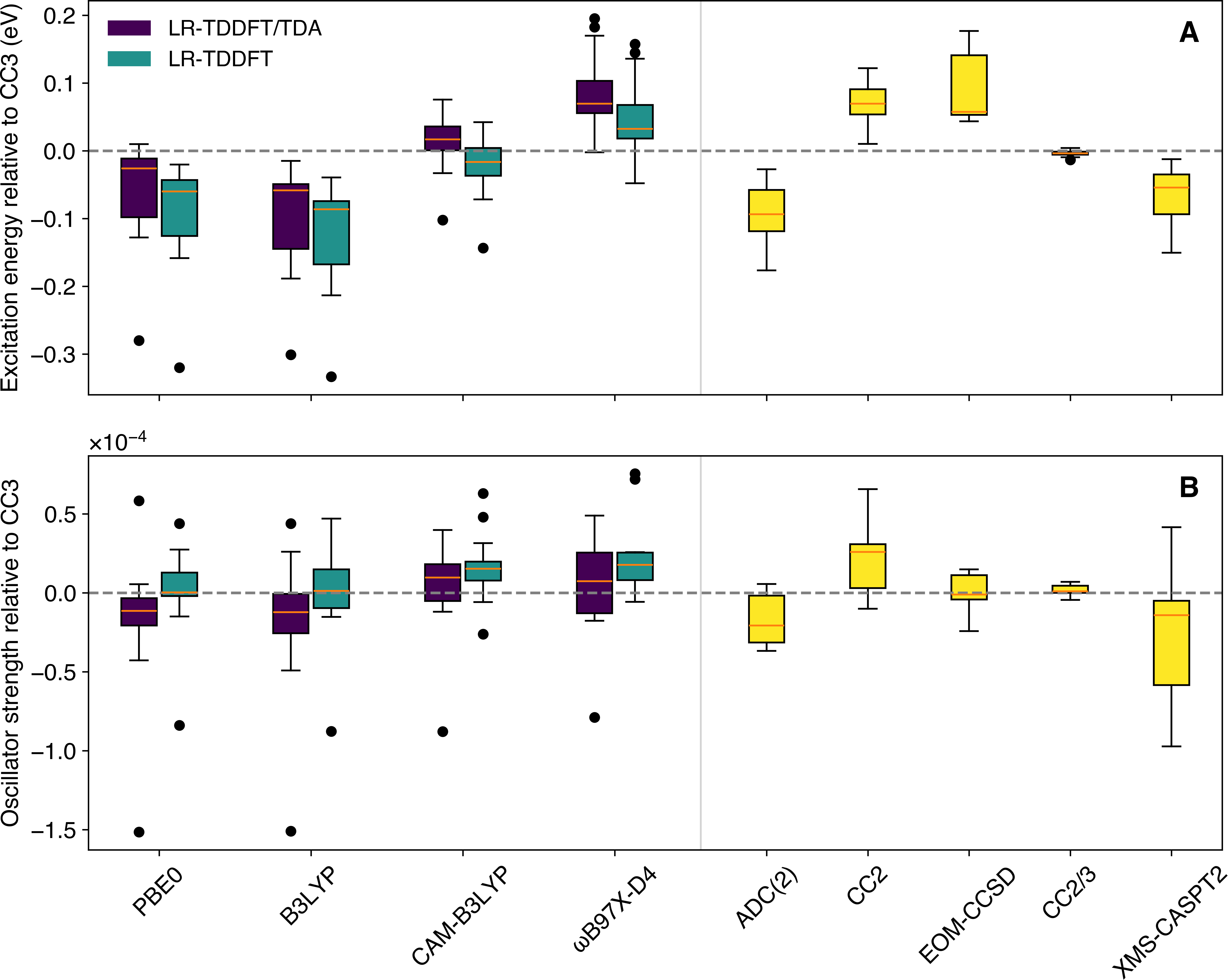}
    \caption{Interquartile range plots of the difference between a range of electronic-structure methods and CC3/aug-cc-pVTZ for the excitation energies (A) and oscillator strengths (B) of the 16 carbonyl-containing molecules presented in Figure \ref{fig:struct}. The central line represents the median, the ends of the box are defined by Q1 (25th percentile; 25\% of the data will be below this) and Q3 (75th percentile; 75\% of the data will be below this), the box size represents the interquartile range (IQR, defined as the distance between Q3 and Q1) and the ends of the whiskers represent the minimum and maximum calculated values (excluding outliers). The outliers are determined using defined upper and lower limits (upper limit = Q3 + 1.5 x IQR; lower limit = Q1 - 1.5 x IQR). If the calculated value falls out of the upper or lower limit, then it is classified as an outlier. Only the lowest-energy transition to S$_1$($n\pi^\ast$) was considered for each molecule. The results for molecules exhibiting a strictly zero oscillator strength for this transition due to symmetry (formaldehyde, acetone, and cyclopropanone) were omitted from the statistics presented in panel B.}
    \label{fig:boxplot}  
\end{figure}

Figure~\ref{fig:boxplot} summarizes the overall results of benchmarking various density-based and wavefunction-based methods for the lowest dark transition of the 16 carbonyl-containing molecules forming our test set (Figure~\ref{fig:struct}). CC3/aug-cc-pVTZ is taken as the reference, and all results are expressed as deviations relative to the reference. Various statistical measures are provided in Table~\ref{tbl:stats} to further characterize the variations between electronic-structure approaches. 

Let us first focus on the results for the excitation energy to the S$_1(n\pi^\ast)$ electronic state of the different carbonyls (Figure~\ref{fig:boxplot}A). From a general perspective, all exchange-correlation functionals tested perform within the expected accuracy for LR-TDDFT(/TDA) for a typical valence transition, namely 0.2-0.3 eV,\cite{laurent2013td} with an extremum MAX value at 0.33 eV for LR-TDDFT/B3LYP. From a trend perspective, the two hybrid functionals assessed, PBE0 and B3LYP, exhibit a small underestimation of the excitation energy in comparison to the reference (negative MSE values) yet with a skewed distribution, while CAM-B3LYP is more balanced (MSE of 0.02 eV with TDA and -0.02 eV without) and $\omega$B97X-D4 slightly overestimates the transition energies. Excitation energies are overall slightly higher with TDA than without for all functionals tested. Long-range corrected functionals, CAM-B3LYP and $\omega$B97X-D4, exhibit less significant outliers than B3LYP and PBE0 (see MAX values in Table~\ref{tbl:stats}). A first striking observation when moving to wavefunction-based methods (Figure~\ref{fig:boxplot}A, right) is the limited number of outliers. ADC(2), with a MSE of -0.09 eV, performs at the same level as LR-TDDFT with hybrid functionals for these dark transitions, while CC2 overestimates their excitation energies (MSE of 0.07 eV), in line with earlier observations for the latter method.\cite{doi:10.1021/acs.jctc.9b01216} Both methods, however, give a smaller SDE than most LR-TDDFT(/TDA) results. Perhaps more surprising is the visible overestimation exhibited by EOM-CCSD transition energies with a MSE of 0.09 eV, yet again with a small SDE of 0.05 eV and a skewed distribution. Overall, the trends in MSE and MAE observed here for the (single-reference) wavefunction-based methods align with the observations of Ref.~\citenum{loos2025questdatabasehighlyaccurateexcitation} for $n\pi^\ast$ states. XMS-CASPT2 provides rather accurate, yet underestimated (MSE$=$-0.06 eV), excitation energies (MAE$=$0.06 eV) with a MAX of 0.15 eV for the excitation energy of glyoxal. Finally, the composite method CC2/3, which is based on a correction of the CC2/aug-cc-pVTZ result with a computationally affordable CC3/def2-SVPD calculation, provides excitation energies in very close agreement with the computationally demanding CC3/aug-cc-pVTZ (with a MAX of 0.01 eV). 

 \begin{table}[h!]
    \caption{Mean signed error (MSE), mean absolute error (MAE), standard deviation of the errors (SDE), and maximum error (MAX) with respect to the CC3/aug-cc-pVTZ reference method for calculated excitation energies (left) and oscillator strengths (right) at the FC point for the 16 carbonyl-containing molecules depicted in Figure~\ref{fig:struct}. The results for molecules exhibiting a strictly zero oscillator strength (formaldehyde, acetone, and cyclopropanone) were omitted from the statistics on oscillator strengths.}
    \label{tbl:stats}
    \begin{tabular}{l cccc|cccc}
    \hline
    \multirow{2}{*}{Method} & \multicolumn{4}{c|}{Excitation energy (eV)} & \multicolumn{4}{c}{Oscillator strength (x10$^{-4}$ a.u)} \\
    {} & MSE & MAE & SDE & MAX & MSE & MAE & SDE & MAX \\
    \hline
    TDA/PBE0   &  -0.06 & 0.06 & 0.07 & 0.28 & -0.19 & 0.30 & 0.47 & 1.51  \\
    PBE0 & -0.09 & 0.09 & 0.07 & 0.32 & 0.00 & 0.18 & 0.30 & 0.84 \\
    TDA/B3LYP & -0.09 & 0.09 & 0.08 & 0.30 & -0.18 & 0.30 & 0.47 & 1.51  \\
    B3LYP & -0.12 & 0.12 & 0.08 & 0.33 & 0.00 & 0.22 & 0.33 & 0.88 \\
    TDA/CAM-B3LYP  & 0.02 & 0.04 & 0.04 & 0.10 & 0.03 & 0.21 & 0.31 & 0.88  \\
    CAM-B3LYP & -0.02 & 0.04 & 0.05 & 0.14 & 0.17 & 0.22 & 0.23 & 0.63 \\
    TDA/$\omega$B97X-D4 & 0.09 & 0.09 & 0.06 & 0.20 & 0.04 & 0.26 & 0.34 & 0.79 \\
    $\omega$B97X-D4 & 0.05 & 0.06 & 0.06 & 0.16 & 0.21 & 0.23 & 0.26 & 0.75 \\
    \hline
    ADC(2) & -0.09 & 0.09 & 0.04 & 0.18 & -0.17 & 0.18 & 0.15 & 0.37 \\
    CC2 & 0.07 & 0.07 & 0.03 & 0.12 & 0.21 & 0.24 & 0.23 & 0.66 \\
    EOM-CCSD & 0.09 & 0.09 & 0.05 & 0.18 & 0.00 & 0.08 & 0.11 & 0.24 \\
    CC2/3 & 0.00 & 0.00 & 0.00 & 0.01 & 0.01 & 0.03 & 0.04 & 0.07 \\
    XMS-CASPT2 & -0.06 & 0.06 & 0.04 & 0.15 & -0.27 & 0.33 & 0.38 & 0.97 \\
    \hline
    \end{tabular}
 \end{table}

We focus now on the performance of each electronic-structure method to capture the small oscillator strength characteristic of dark transitions (Figure~\ref{fig:boxplot}B, note the $\times 10^{-4}$ on top of the ordinate axis). The reader should bear in mind that the results presented here are averaged over molecules exhibiting transitions with an oscillator strength magnitude that can significantly differ, with the smallest being $1.95\times 10^{-6}$ (cyclobutanone) and the largest being $2.92\times 10^{-4}$ (glyoxal) according to the CC3/aug-cc-pVTZ. We offer a different representation of Figure~\ref{fig:boxplot}B in the SI that accounts for the absolute value of each oscillator strength. We also note that we omitted from the statistics presented here the results for molecules exhibiting a strictly zero oscillator strength due to symmetry (formaldehyde, acetone, and cyclopropanone). Several outliers (among which glyoxal, known for its rather challenging electronic structure given the conjugated nature of its two carbonyl moieties\cite{glyoxalelectronicstructure}) are observed for the oscillator strengths obtained with LR-TDDFT(/TDA) in comparison to the wavefunction-based methods, and LR-TDDFT(/TDA) exhibits overall larger SDE and MAX values (Table~\ref{tbl:stats}). Oscillator strengths obtained within TDA are worse in quality when using hybrid functionals (compare MAE and SDE values in Table~\ref{tbl:stats}), while this difference reduces slightly when moving to long-range corrected functionals. The balanced performance of CAM-B3LYP for oscillator strengths aligns with the conclusion of earlier works.\cite{doi:10.1021/acs.jctc.0c01228,caricato2010oscillator} As highlighted above, wavefunction-based methods exhibit fewer outliers overall. ADC(2) appears to slightly underestimate oscillator strengths (MSE$=-0.17\times 10^{-4}$) and CC2 overestimate them (MSE$=0.21\times 10^{-4}$). EOM-CCSD provides accurate oscillator strengths (MAE$=0.08\times 10^{-4}$ and SDE$=0.11\times 10^{-4}$). The composite approach CC2/3 again manages to correct the CC2 results to offer oscillator strengths in closest agreement with the reference. Mixed results are obtained with XMS-CASPT2. Overall, the method exhibits the largest MSE with $-0.27\times 10^{-4}$, with an SDE close to that obtained with LR-TDDFT(/TDA) using hybrid functionals. Earlier works reported that XMS-CASPT2 underestimates oscillator strengths for dark transitions, in line with the MSE obtained here.\cite{Prljphotolysis2020,prlj2023deciphering} We also note that the oscillator strengths obtained with XMS-CASPT2 depend on the magnitude of the (imaginary) shift. For example, the oscillator strengths characterizing the transition to the S$_1(n\pi^\ast)$ state for acetaldehyde is $4.19\times 10^{-5}$ with an imaginary shift of 0.1 a.u., $3.60\times 10^{-5}$ with an imaginary shift of 0.3 a.u., and $2.80\times 10^{-5}$ with an imaginary shift of 0.5 a.u. We also calculated the same oscillator strength (acetaldehyde) with OpenMolcas and obtained a value of $3.15\times 10^{-6}$, to be compared with $4.19\times 10^{-5}$ obtained with Bagel (the same basis set, active space, SS-SR strategy, and imaginary shift were employed). This order of magnitude difference can be understood by the strategy used by OpenMolcas to determine oscillator strengths, where oscillator strengths are obtained by combining SA-CASSCF transition dipole moments with XMS-CASPT2 energy difference between the electronic states considered. (We also note that Bagel uses a density-fitting approach to further speed up the XMS-CASPT2 calculations.) This approach may work for transitions with a large oscillator strength but should be used with care for small oscillator strengths.

\subsection{Benchmark away from the Franck-Condon point}
\label{sec:beyondFCP}

\begin{figure}[H]
    \centering
    \includegraphics[width=0.65\textwidth]{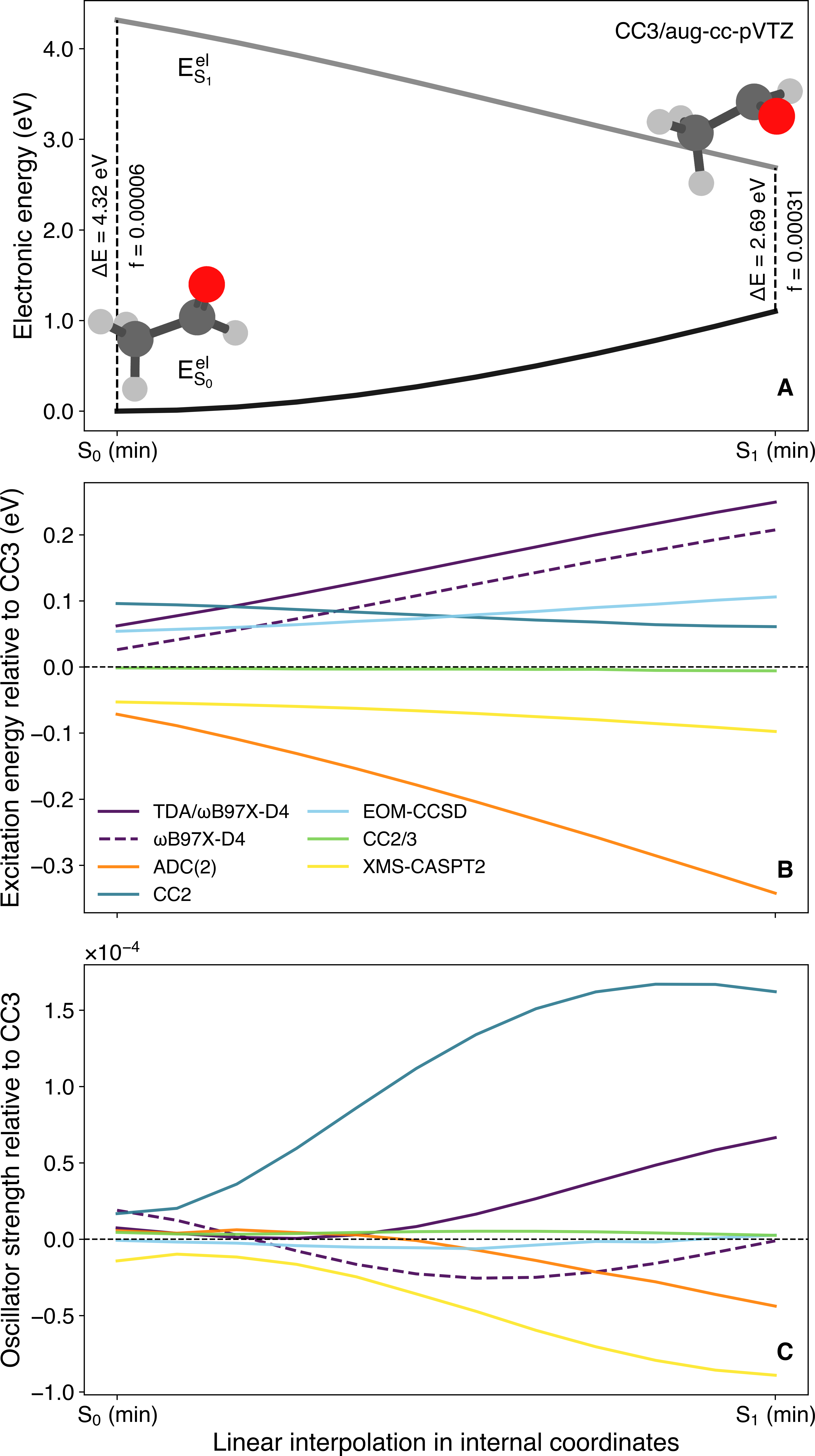}
    \caption{(A) Linear interpolation in internal coordinates (LIIC) connecting the ground-state optimized geometry, S$_0$ (min), to the first-excited state optimized geometry, S$_1$ (min) with a $n\pi^\ast$ character, for acetaldehyde.  Electronic energies (and oscillator strengths) were obtained at the CC3/aug-cc-pVTZ level of theory. The critical molecular geometries of acetaldehyde, S$_0$ (min) and S$_1$ (min), are given as insets.
    Excitation energies (B) and oscillator strengths (C) between S$_1$ (min) and S$_0$ (min) along the LIIC pathway, represented as a deviation relative to the CC3/aug-cc-pVTZ reference values (horizontal dashed line).}
    \label{fig:liic}
\end{figure}

We now move our attention to the accuracy of electronic-structure methods in describing dark transitions \textit{away} from the FC point. To achieve this goal, we used acetaldehyde as a representative carbonyl-containing molecule and located its minimum-energy geometry in the S$_1(n\pi^\ast$) electronic state using CC2/cc-pVTZ. A LIIC path was then created to obtain geometries smoothly connecting the S$_0$ minimum-energy geometry to the S$_1$ minimum-energy geometry. Electronic energies and oscillator strengths can then be calculated on the support of the LIIC geometries with each electronic-structure method tested throughout this work. The resulting LIIC pathway obtained with the reference method (CC3/aug-cc-pVTZ) is depicted in Figure~\ref{fig:liic}A and shows the smooth decrease in S$_1$ electronic energy when moving from S$_0$ (min) -- the FC point -- to S$_1$ (min). The ground-state energy rises along this curve, due to the distortion away from the FC point -- see the molecular representations depicting the S$_0$ (min) and S$_1$ (min) geometries. It is important to remark here that the S$_1$ (min) geometry exhibits only slight distortions in comparison to that of S$_0$ (min) -- the methyl group rotated and the carbon of the carbonyl group acquired a small pyramidalization. Yet, this minor distortion is sufficient to reduce the energy gap between S$_1$ and S$_0$ from 4.32 eV at S$_0$ (min) to 2.69 eV at S$_1$ (min). Perhaps more important is to realize that the oscillator strength characterizing the S$_0$ to S$_1(n\pi^\ast$) transition changes by an order of magnitude along the pathway, from $6\times 10^{-5}$ at S$_0$ (min) to $3\times 10^{-4}$, highlighing the impact of non-Condon effects on this dark transition. In the following, we compare the excitation energies and oscillator strengths obtained along this LIIC pathway for acetaldehyde, using the CC3/aug-cc-pVTZ as a reference (Figure~\ref{fig:liic}B and C). In this representation, a method consistently agreeing with the reference along the path would be characterized by a mostly horizontal line (the shift from the horizontal line at 0.0 eV being a constant shift away from the reference), whereas any result characterized by a non-linear curve indicates a deterioration (or improvement) of the quality of the method along the LIIC pathway. 

Let us begin by comparing the performance of electronic-structure methods in describing the excitation energy (or electronic-energy gap) between S$_0$ and S$_1(n\pi^\ast)$ along the LIIC pathway (Figure~\ref{fig:liic}B). The LR-TDDFT and LR-TDDFT/TDA excitation energies, here calculated using the $\omega$B97X-D4 functional (the results obtained with other functionals, presented in the SI, follow similar trends), exhibit a monotonically linear increase when evolving towards the S$_1$ (min) geometry. TDA does not alter this behaviour but further enhances the positive energy shift in comparison to the reference (mirroring our earlier observations on the performance of this functional in Figure~\ref{fig:boxplot}). The excitation energies obtained with ADC(2) show a rapid deviation from the reference when progressing towards S$_1$ (min), reaching the largest deviation from the reference upon reaching this point. This behaviour is in line with a recently characterized issue with ADC(2) when describing electronic states with $n\pi^\ast$ character,\cite{marsili2021caveat} where a too shallow S$_1(n\pi^\ast)$ potential energy surface in ADC(2)\cite{hattig2005structure,budzak2017benchmarkadc2} combined with a MP2 reference becoming invalid upon distortion of the carbonyl group leads to artificially small energy gaps between S$_0$ and S$_1(n\pi^\ast)$.\cite{marsili2021caveat} Excitation energies obtained with CC2 follow the reference closely, exhibiting a small overestimation. Correcting the CC2 results with the composite method CC2/3 leads to results nearly indistinguishable from the reference (dashed horizontal line). The excitation energies obtained with EOM-CCSD along the LIIC pathway mirror closely those obtained with CC2, yet with a minute monotonic increase in deviation from the reference (instead of a small decrease for CC2). XMS-CASPT2 offers excitation energies of constant quality in comparison to the reference, with only a small underestimation of the gap (a trend echoing our earlier observations based on Figure~\ref{fig:boxplot}). 

We turn now to the oscillator strengths along the LIIC and their deviation from the reference (Figure~\ref{fig:liic}C). We can note first that EOM-CCSD and the composite method CC2/3 offer oscillator strengths in close agreement with the reference along the LIIC pathway. The performance of the composite method CC2/3 was not a given if one contrasts its results for oscillator strength with those obtained from CC2. The oscillator strengths obtained with CC2 for the first part of the LIIC, close to S$_0$ (min), follow those of the reference, but an abrupt increase is then observed for CC2. Recalling that the oscillator strength of acetaldehyde grows by an order of magnitude between S$_0$ (min) and S$_1$ (min) (see discussion at the beginning of this Section), the oscillator strength obtained with CC2 near the S$_1$ (min) deviates by nearly 50\% from the reference. ADC(2) provides oscillator strengths in overall close agreement with the reference (besides a small decrease in accuracy when reaching the region of the S$_1$ minimum), in stark contrast with its mediocre performance for excitation energies. LR-TDDFT and LR-TDDFT/TDA exhibit quite different behaviors when it comes to oscillator strengths: LR-TDDFT/$\omega$B97X-D4 oscillates around the reference along the LIIC pathway, whereas LR-TDDFT/TDA/$\omega$B97X-D4 shows a more stable agreement with the reference for the first part of the LIIC pathway, before exhibiting a monotonic increase when getting close to the S$_1$ (min). XMS-CASPT2 predicts oscillator strengths deviating from the reference when leaving the FC region, producing smaller values as expected from the results obtained in Section~\ref{sec:FCP}.

\subsection{Benchmark around the Franck-Condon point}
\label{sec:FAP}

\begin{figure}[h!]
    \centering
    \includegraphics[width=1.0\textwidth]{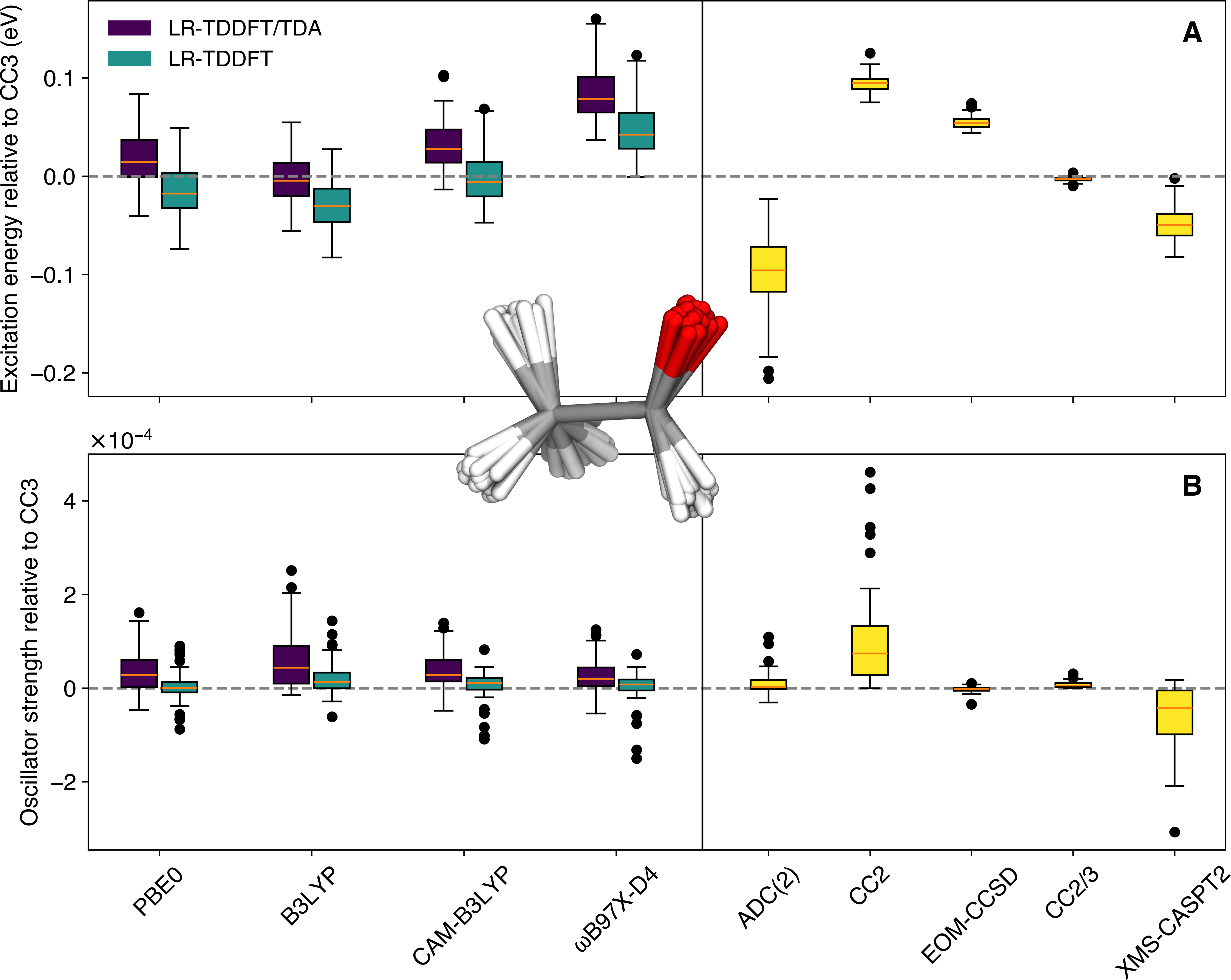}
    \caption{Interquartile range plots of the difference between a range of electronic-structure methods and CC3/aug-cc-pVTZ for the excitation energies (A) and oscillator strengths (B) obtained for 50 geometries sampled from the harmonic Wigner distribution of acetaldehyde. An overlay of the sampled geometries is provided as an inset. See the caption of Figure~\ref{fig:boxplot} for a definition of the ranges presented. Only the lowest-energy transition to S$_1$($n\pi^\ast$) was considered for each molecule. }
    \label{fig:boxplotNEA}  
\end{figure}

In this last Section, we propose to investigate the consistency of the various electronic-structure methods tested in predicting excitation energies and oscillator strengths in the vicinity of the FC point, i.e., not as far in the nuclear configuration space as the S$_1$ minimum tested in Section~\ref{sec:beyondFCP}. To achieve this goal, we approximated the ground-state distribution of acetaldehyde at 0 K by using a harmonic Wigner distribution, from which we sampled 50 optimal geometries. Excitation energies and oscillator strengths are calculated on the support of these geometries, and the result is used to depict the photoabsorption cross-section of the molecule following the NEA (see Section~\ref{sec:method} for additional details). These molecular geometries are part of the FC region but exhibit small distortions away from the FC point, as depicted in the inset of Figure~\ref{fig:boxplotNEA}, allowing us to further test the accuracy of the selected electronic-structure methods in capturing non-Condon effects. We can then compare the calculated photoabsorption cross-sections for acetaldehyde and discuss how the choice of the electronic-structure method can impact its predicted photolysis half-life ($t_{1/2}=\frac{\ln(2)}{j}$).

The general trends observed for the excitation energies obtained by the different electronic-structure methods tested relative to the reference for the 50 sampled geometries of acetaldehyde (Figure~\ref{fig:boxplotNEA}A) align with those observed earlier for the test set of 16 carbonyl-containing molecules (Figure~\ref{fig:boxplot}A). LR-TDDFT/TDA performs slightly better than LR-TDDFT for hybrid functionals, and CAM-B3LYP offers a more balanced description of excitation energies than $\omega$B97X-D4, both with and without TDA. Overall, both LR-TDDFT and LR-TDDFT/TDA provide a rather consistent and accurate representation of the excitation energies for the 50 sampled geometries of acetaldehyde, with all functionals tested. As expected from our earlier findings, ADC(2) on average underestimates the excitation energies, whereas CC2 slightly overestimates them, but with a small standard deviation in comparison to ADC(2). Even if slightly overestimated, the excitation energies obtained with EOM-CCSD show a high degree of similarity with the reference. The composite method CC2/3, again, offers excitation energies in near-perfect agreement with the CC3/aug-cc-pVTZ result. The excitation energies obtained with XMS-CASPT2 are only slightly underestimated but overall exhibit a rather small spread.

Comparing now the oscillator strengths obtained with LR-TDDFT(/TDA) for the 50 representative geometries of acetaldehyde to the reference (Figure~\ref{fig:boxplotNEA}B), we observe a consistent result for all functionals tested, but with a marked standard deviation (in particular for LR-TDDFT/TDA) and a significant number of outliers. As observed for our test beyond the FC point (Section~\ref{sec:beyondFCP}), ADC(2) and CC2 have an opposite behavior when it comes to their accuracy to predict excitation energies and oscillator strengths. While the excitation energies obtained with ADC(2) showed the largest deviation among all methods tested (Figure~\ref{fig:boxplotNEA}A), its oscillator strengths are in close agreement with the reference. Conversely, the good performance of CC2 for excitation energy is somewhat contrasted by its poor description of oscillator strengths. More specifically, the oscillator strengths obtained with CC2 are larger than those of the reference, a behavior in line with that observed for the CC2 oscillator strength beyond the FC point (see Figure~\ref{fig:liic}), and the CC2 outliers exhibit geometrical distortions (in particular, out-of-plane bending around the carbonyl moiety) resembling those observed in the LIIC when approaching S$_1$ (min) (see Section~\ref{sec:beyondFCP}).  As expected from the previous Section, the oscillator strengths predicted by EOM-CCSD and CC2/3 are in close agreement with the reference and exhibit only a minute standard deviation. XMS-CASPT2 provides oscillator strengths that are overall smaller than those obtained with the reference, again confirming the trends observed in the two previous Sections.

\begin{figure}[h!]
    \centering
    \includegraphics[width=0.9\textwidth]{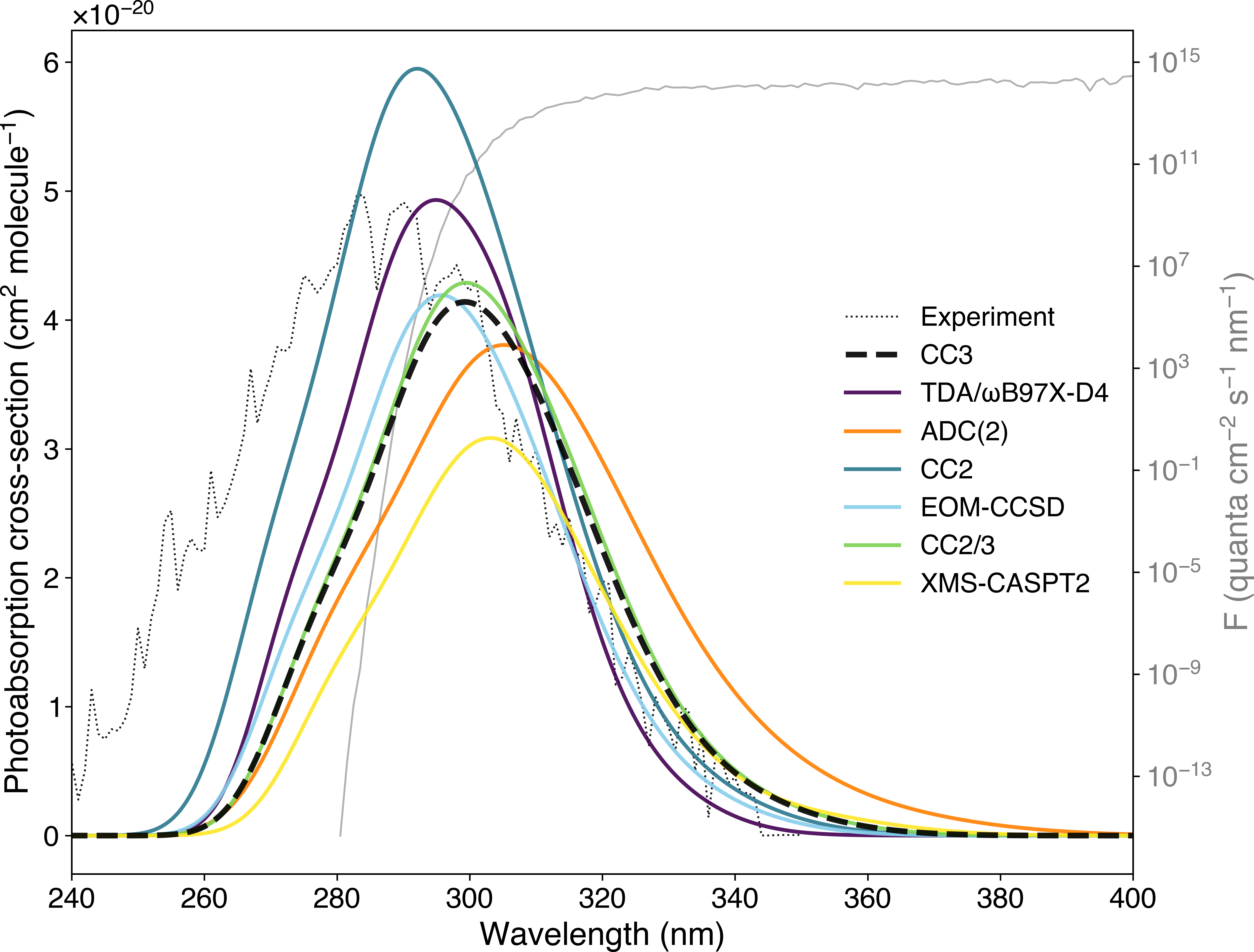}
    \caption{Photoabsorption cross-section of acetaldehyde obtained with the NEA using different electronic-structure methods (only the lowest-energy transition S$_0\rightarrow$S$_1(n\pi^\ast)$ was considered). All photoabsorption cross-sections presented were obtained from the same set of 50 optimally-sampled geometries, from which excitation energies and oscillator strengths were calculated at a given level of electronic-structure theory. The experimental photoabsorption cross-section was presented in Ref.~\citenum{limaovieira2003acetalcross} and obtained via the MPI-Mainz UV/Vis Spectral Atlas.\cite{keller2013mpi} The standardized medium actinic flux $F(\lambda)$ (solar zenith angle = 60$^\circ$, overhead ozone column = 350 DU) for a ground elevation of 0 km above sea level is represented in a log scale. 
    }
    \label{fig:crosssection}  
\end{figure}

The predicted photoabsorption cross-section of acetaldehyde can be determined for each electronic-structure method using the NEA, based on the excitation energies and oscillator strengths calculated on the support of the 50 representative geometries. Figure~\ref{fig:crosssection} presents the photoabsorption cross-sections obtained with the different methods tested in this work and the CC3 reference, together with the experimental photoabsorption cross-section.\cite{limaovieira2003acetalcross} Let us first address the main surprising result: an apparent discrepancy between the overall width of the photoabsorption cross-section obtained with CC3, our reference throughout this work, and that of the experimental photoabsorption cross-section. This difference should not come as a surprise, though: the NEA, which is a numerical realization of the reflection principle,\cite{schinkebook} does not describe vibronic progressions resulting from the overlap between the vibrational wavefunction of the ground electronic states and the vibrational wavefunctions of the excited state(s).\cite{crespo2014spectrum,crespo2018recent,prlj2021calculating} Accounting for vibronic progressions would require the calculation of Franck-Condon (and Herzberg-Teller) factors or the use of quantum dynamics simulations.\cite{santoro2016beyondvertical,prlj2021calculating} Hence, the NEA produces photoabsorption cross-sections in quantitative agreement with experiment for transitions involving dissociative states (as long as the electronic-structure method used describes the states of interest adequately).\cite{C8CP00199E,prlj2021calculating,prlj2023deciphering} The NEA may suffer variably from its approximations for photoabsorption cross-sections involving transitions to bound states,\cite{crespo2014spectrum,crespo2018recent,srsen2020nealimits,prlj2021calculating} as observed here for the S$_0\rightarrow$S$_1(n\pi^\ast)$ of acetaldehyde. We compare the photoabsorption cross-section calculated with the NEA to that obtained with a Franck-Condon Herzberg-Teller formalism using LR-TDDFT/TDA in the SI (Fig.~S10), confirming the expected shift in energy and shape between NEA and vibronically-resolved spectra. This paragraph highlights the challenge to compare the results of electronic-structure methods directly to experiment for observables like photoabsorption cross-sections.

Despite the narrower photoabsorption cross-sections produced by the NEA, all electronic-structure methods predict absolute cross-section values in good agreement with the maximum observed experimentally, i.e., $\sigma(\lambda_{\text{max}}=284)=4.97\times10^{-20} \text{cm}^2\text{molecule}^{-1}$. It is instructive to measure the impact of the deviations in excitation energies and oscillator strengths discussed above (when analyzing Figure~\ref{fig:boxplotNEA}) on the produced photoabsorption cross-sections. Hence, the photoabsorption cross-section obtained with CC2 exhibits the largest deviation from the reference in terms of intensity and blueshift of the band. The smaller oscillator strengths obtained with XMS-CASPT2 result in a photoabsorption cross-section lower in intensity than the reference (as observed previously with this method), yet with a band centered near the CC3 photoabsorption cross-section. The rather large spread in excitation energies obtained with ADC(2) results in a photoabsorption cross-section that appears wider than the reference and the other calculated cross-sections, with a tail reaching longer wavelengths. CC2/3 and EOM-CCSD produced a photoabsorption cross-section in close agreement with the reference, while the slightly larger oscillator strengths and excitation energies of LR-TDDFT/TDA/$\omega$B97X-D4 lead to a cross-section close to the CC3 reference but with a minute shift to higher energy and intensity. 

As a final sensitivity analysis, we propose to calculate the resulting photolysis half-life of acetaldehyde obtained by using the predicted photoabsorption cross-section for each electronic-structure method based on Eq.~\eqref{eq:photolysis}. We considered a quantum yield $\phi(\lambda)\approx 1$ and a typical medium flux for the actinic flux $F(\lambda)$ (see Section~\ref{sec:method} for additional details and Figure~\ref{fig:crosssection} for a depiction of $F(\lambda)$ using a log scale). With the photoabsorption cross-section predicted by CC3, acetaldehyde would exhibit a half-life of 5.8 hours upon photolysis. The minute variation of the photoabsorption cross-sections between CC3 and EOM-CCSD is enough to extend the photolysis half-life of acetaldehyde to 8.8 hours, as the $\sim$0.05 eV shift towards shorter wavelengths displayed by the EOM-CCSD photoabsorption cross-section reduces its overlap with the actinic flux. The predicted photoabsorption cross-sections of CC2 and LR-TDDFT/TDA/$\omega$B97X-D4, further shifted to higher energy than that of EOM-CCSD, lead to a half-life upon photolysis of 7.1 hours and 11.0 hours, respectively. The tail in the low-energy part of the spectrum exhibited by the photoabsorption cross-section obtained with ADC(2) results in a large overlap with the actinic flux, resulting in a shorter predicted half-life of 3.0 hours.

\section{Conclusions}

In summary, our work offered an overview of the performance of the most employed electronic-structure methods to describe dark, $n\pi^\ast$ transitions of carbonyl-containing molecules, of great importance for atmospheric chemistry.

Using a test set composed of 16 compounds and CC3 as a reference, we tested the performance of LR-TDDFT(/TDA), ADC(2), CC2, EOM-CCSD, XMS-CASPT2, and the composite method CC2/3 in describing the excitation energy and oscillator strength for the low-lying $n\pi^\ast$ transition of these molecules. The results for excitation energies are in line with the performance of these methods for typical valence transitions, with LR-TDDFT(/TDA) showing a rather large standard deviation with most functionals tested. ADC(2) appears to underestimate the energy of this type of transition, while CC2 and EOM-CCSD slightly overestimate it. Regarding the oscillator strengths, LR-TDDFT(/TDA) also exhibits a rather large standard deviation for the compounds studied. Wavefunction-based methods were more accurate, with a trend for ADC(2) to underestimate the intensity of this transition, while CC2 may overestimate it.

Dark transitions are known to pick up intensity \textit{via} nuclear displacement. Hence, we tested whether the performance observed at the FC point for all methods would degrade when displacing the molecular geometry, here of acetaldehyde, beyond the FC point towards the S$_1$ minimum of this molecule (still exhibiting $n\pi^\ast$ character). For excitation energies, ADC(2) and LR-TDDFT(/TDA) exhibited an increasing drift away from the reference when leaving the FC point. For oscillator strengths, the result of CC2 deteriorated the further away the molecular geometry of acetaldehyde was moved from the FC point. This potential inhomogeneity of the performance of each method for various regions of nuclear configuration space highlights the critical importance of testing electronic-structure methods beyond the FC point.

To measure the impact of the deviations observed at and beyond the FC point, we also sampled 50 molecular geometries representative of the ground-state distribution of acetaldehyde and compared the performance of the electronic-structure methods for excitation energies and oscillator strengths, but also their predicted photoabsorption cross-section. The results obtained confirmed the observations at and beyond the FC point. The performance of LR-TDDFT(/TDA) depends on the functional employed, and LR-TDDFT/TDA performs slightly worse than LR-TDDFT. CC2 tends to overestimate the energy and the oscillator strength for this type of transition, whereas ADC(2) offers rather accurate oscillator strengths but underestimates the excitation energy. To provide a 'real-world' assessment of the impact of the observed deviations between electronic-structure methods, we also calculated the photolysis half-life of acetaldehyde as predicted by each method. With a reference value of 5.8 hours when using the CC3 photoabsorption cross-section, the calculated photolysis half-life varies from 3.0 hours with the photoabsorption cross-section obtained by ADC(2) to 11.0 hours with that predicted by LR-TDDFT/TDA/$\omega$B97X-D4.

Throughout all our tests, the composite method CC2/3 always accurately reproduced the reference CC3 results. This achievement is remarkable given the rather large deviations for oscillator strengths obtained with CC2 away from the FC point, and places this composite method as an ideal compromise to CC3 for larger molecular systems. XMS-CASPT2 provided excitation energies in good agreement with the reference, at and beyond the FC point. With the active space employed here and in line with previous works, XMS-CASPT2 appears to underestimate the oscillator strengths of dark transitions. 

Our work provides an attempt to develop strategies based on theoretical best estimates to benchmark electronic-structure methods \textit{beyond} the FC point -- of particular importance for probing dark states such as the $n\pi^\ast$ transitions of carbonyls. The results presented here can hopefully offer guidance for future works on the photochemistry of carbonyl-containing molecules, more specifically in atmospheric chemistry, and stimulate further benchmarking of electronic-structure methods, including our reference CC3, for excitation energies and other electronic properties beyond the FC point.

\begin{acknowledgement}

We would like to thank Štěpán Sršeň for technical help with the PyNEAppLES code and useful advice regarding the optimal sampling approach.
This project has received funding from the European Research Council (ERC) under the European Union's Horizon 2020 research and innovation programme (Grant agreement No. 803718, project SINDAM) and the EPSRC Grants EP/V026690/1, EP/Y01930X/1, and EP/X026973/1. This work was carried out using the computational facilities of the Advanced Computing Research Centre, University of Bristol -- \url{http://www.bristol.ac.uk/acrc/}. The authors acknowledge the use of resources provided by the Isambard 3 Tier-2 HPC Facility. Isambard 3 is hosted by the University of Bristol and operated by the GW4 Alliance (\url{https://gw4.ac.uk}) and is funded by UK Research and Innovation; and the Engineering and Physical Sciences Research Council [EP/X039137/1].

\end{acknowledgement}

\begin{suppinfo}

Additional computational details (test of basis sets, convergence of CC2/3), plots with normalized oscillator strengths, additional test of exchange-correlation functionals along the LIIC pathway, and depictions of the active-space orbitals. (PDF) \\
The data generated by this work (excitation energies, oscillator strengths, geometries, scripts) is provided in a zip file. (ZIP)

\end{suppinfo}

\providecommand{\latin}[1]{#1}
\makeatletter
\providecommand{\doi}
  {\begingroup\let\do\@makeother\dospecials
  \catcode`\{=1 \catcode`\}=2 \doi@aux}
\providecommand{\doi@aux}[1]{\endgroup\texttt{#1}}
\makeatother
\providecommand*\mcitethebibliography{\thebibliography}
\csname @ifundefined\endcsname{endmcitethebibliography}
  {\let\endmcitethebibliography\endthebibliography}{}

\end{document}


\tableofcontents

\section{Test of the basis sets}

\begin{figure}[ht!]
     \centering
     \includegraphics[width=0.95\textwidth]{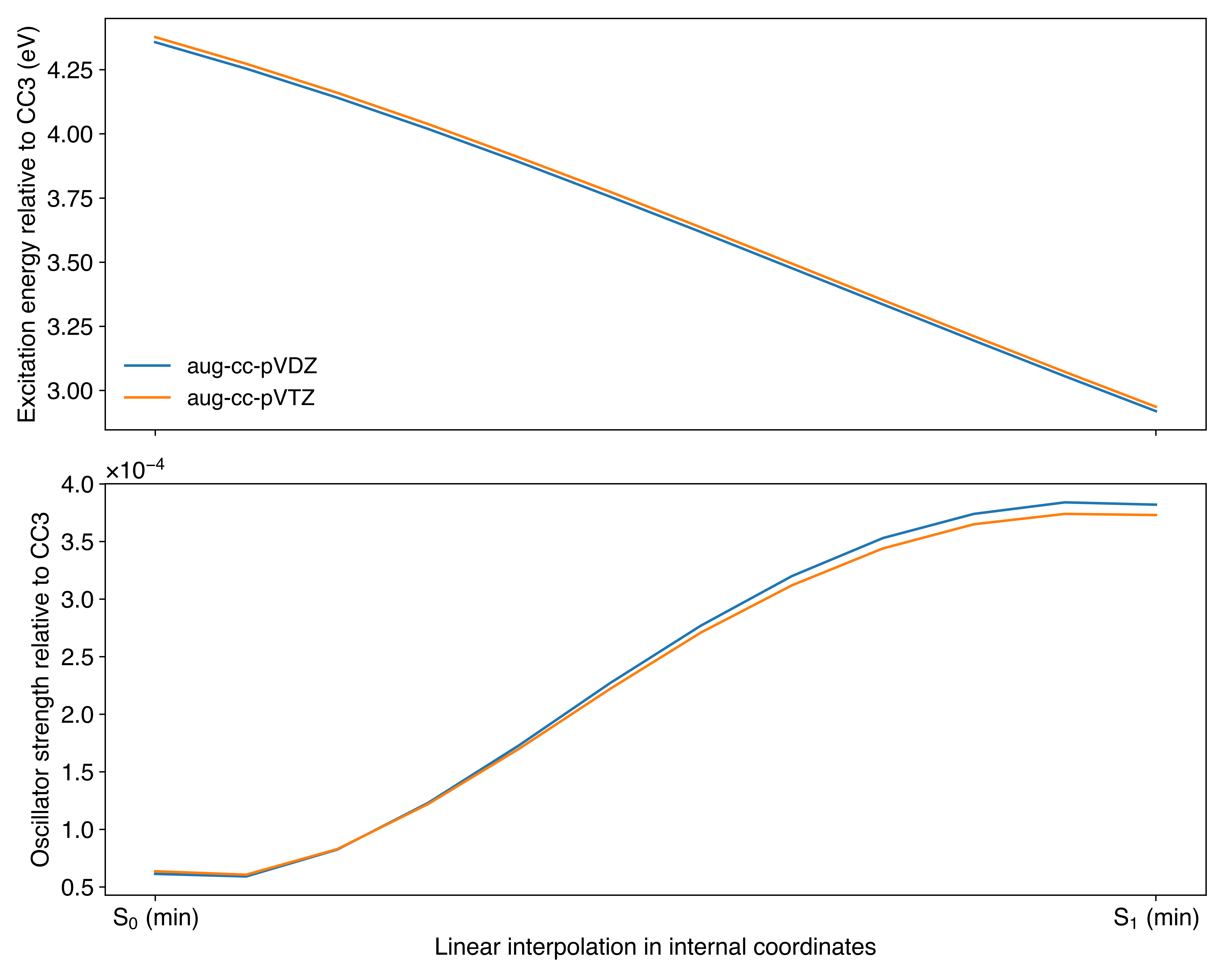}
     \caption{LIIC pathways comparing the impact of the basis set for excitation energies and oscillator strengths of acetaldehyde ($n\pi^\ast$ transition), calculated using LR-TDDFT/TDA/$\omega$B97X-D4 and either aug-cc-pVDZ or aug-cc-pVTZ.}
     \label{fig:si-liic-basis}
 \end{figure}

Table \ref{tbl:basis-set} shows the importance of diffuse functions to describe the oscillator strength for a $n\pi^\ast$ transition (here for acetaldehyde). Additional tests of the basis set away from the Franck-Condon point (using the LIIC described in the main text) revealed a small yet sizeable deviation for the oscillator strength obtained with aug-cc-pVDZ and aug-cc-pVTZ when reaching the region of S$_1$ (min) -- see Figure \ref{fig:si-liic-basis} -- motivating us to use aug-cc-pVTZ throughout this work for all electronic-structure methods tested. 

 \begin{table}[ht!]
    \caption{Vertical excitation energies ($\Delta E^{\text{el}}$) and oscillator strengths ($f$) for the lowest singlet transition ($n\pi^\ast$) of acetaldehyde, calculated using LR-TDDFT/TDA with PBE0 and EOM-CCSD across various Dunning's basis sets.}
    \label{tbl:basis-set}
    \begin{tabular}{l|cc|cc}
    \hline
    \multirow{2}{*}{} & \multicolumn{2}{c|}{LR-TDDFT/TDA/PBE0} & \multicolumn{2}{c}{EOM-CCSD} \\
    & \multicolumn{1}{l}{$\Delta E^{\text{el}}$ (eV)} & \multicolumn{1}{l|}{$f$ (x10$^{-5}$ a.u.)} & \multicolumn{1}{l}{$\Delta E^{\text{el}}$ (eV)} & \multicolumn{1}{l}{$f$ (x10$^{-5}$ a.u.)} \\ \hline
    cc-pVDZ & 4.34 & 0.49 & 4.43 & 1.81 \\
    cc-pVTZ & 4.35 & 2.07 & 4.41 & 3.25 \\
    cc-pVQZ & 4.34 & 2.68 & - & - \\
    aug-cc-pVDZ & 4.30 & 3.99 & 4.37 & 5.39 \\
    aug-cc-pVTZ & 4.32 & 4.47 & 4.37 & 5.54 \\
    aug-cc-pVQZ & 4.31 & 4.25 & - & - \\ \hline
    \end{tabular}
\end{table}

\section{Impact of the small basis set selected for the CC2/3 calculations}

 \begin{figure}[H]
     \centering
     \includegraphics[width=0.95\textwidth]{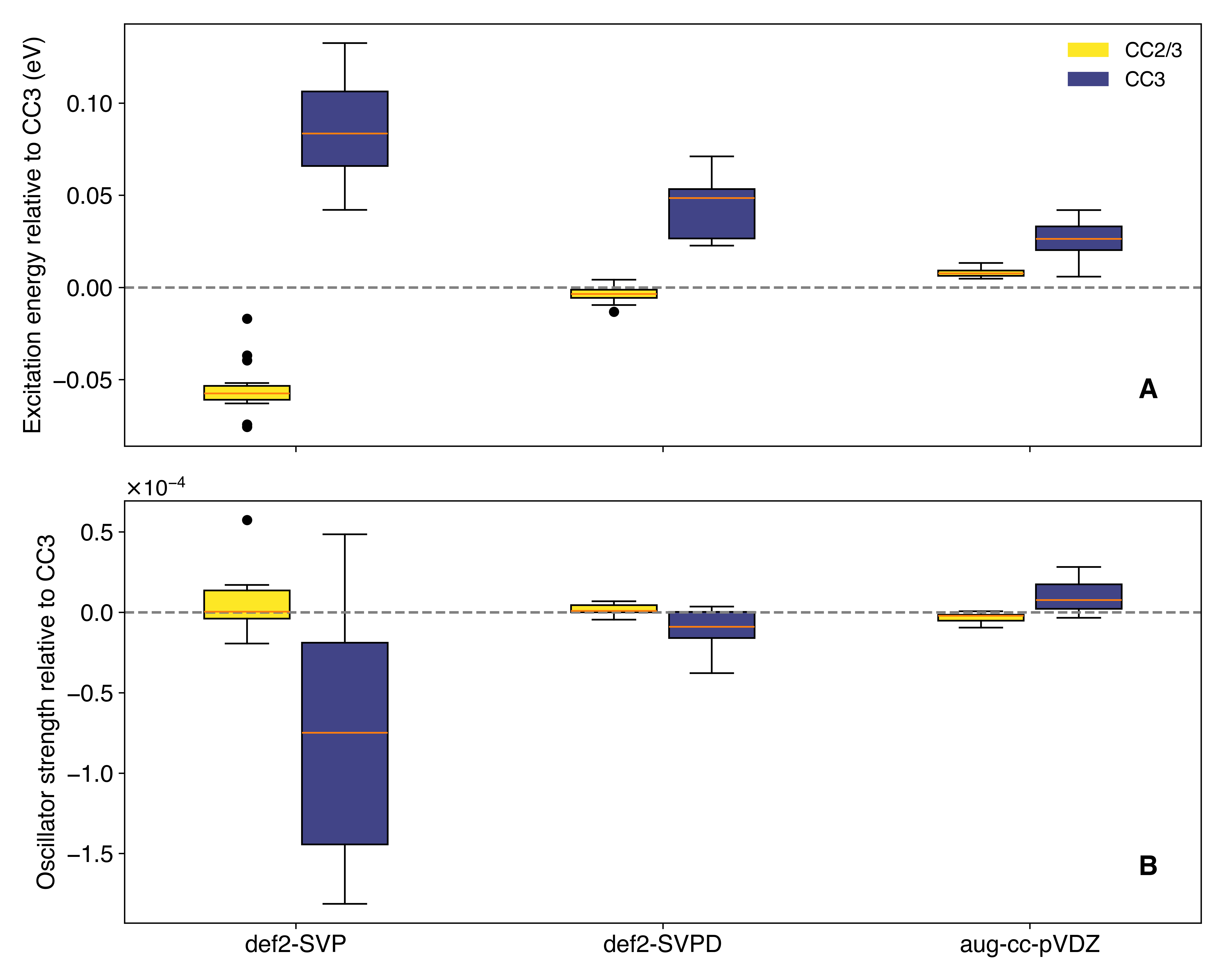}
     \caption{Interquartile range plots of the difference between CC2/3 and CC3 methods with a range of basis sets versus CC3/aug-cc-pVTZ for the excitation energies (A) and oscillator strengths (B) of the 16 carbonyl-containing molecules presented in the main text. The basis sets indicated are those used for the 'low-cost' CC3 calculation, showing that diffuse functions are important to reach the full accuracy of the composite method for the $n\pi^\ast$ transition studied. The results for molecules exhibiting a strictly zero oscillator strength for the lowest $n\pi^\ast$ transition due to symmetry (formaldehyde, acetone, and cyclopropanone) were omitted from the statistics presented in panel B.}
     \label{fig:si-cc2-3}
 \end{figure}

\section{Normalized deviations for the oscillator strengths}

 \begin{figure}[H]
     \centering
     \includegraphics[width=0.95\textwidth]{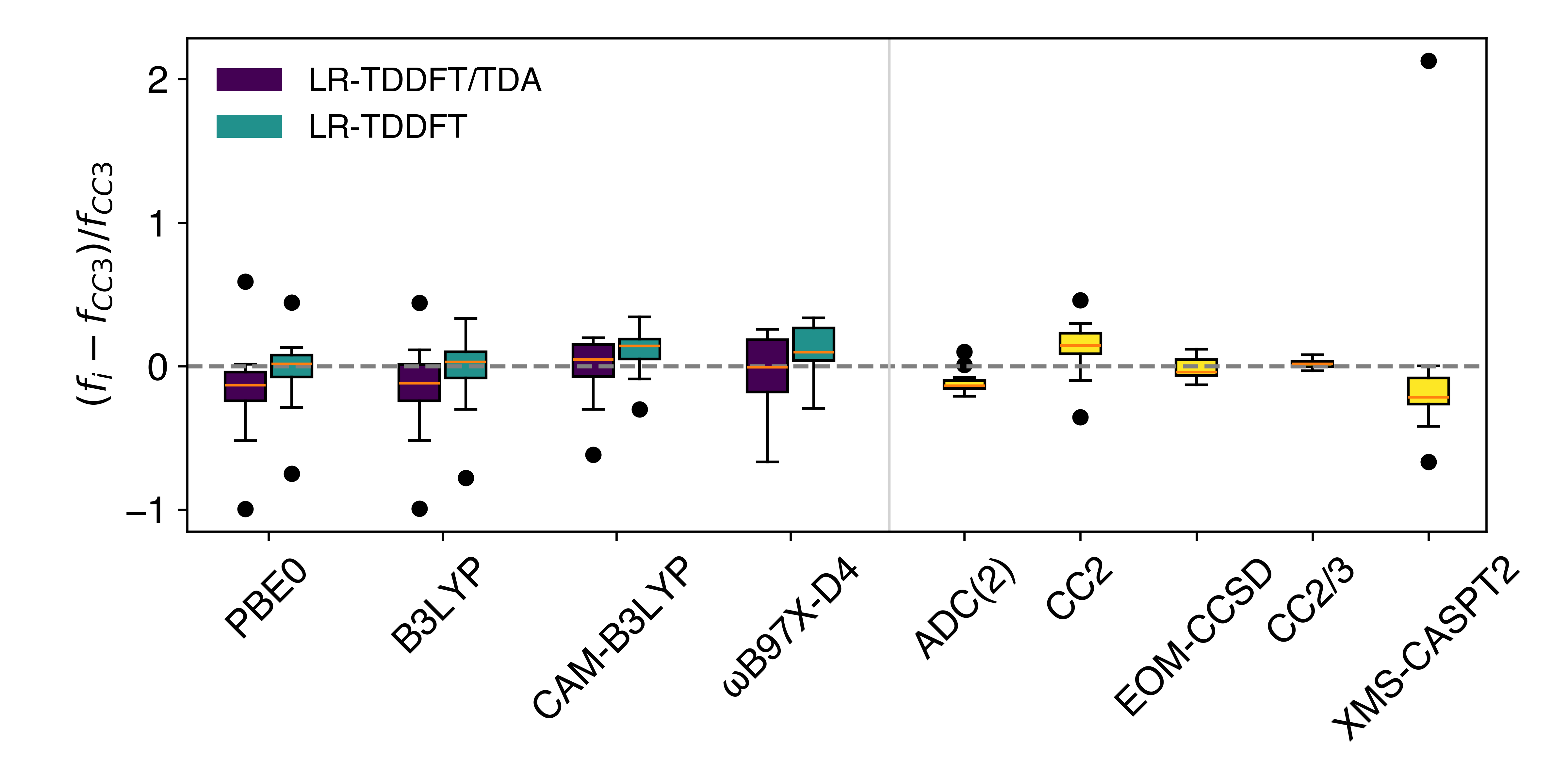}
     \caption{Interquartile range plots of the difference between a range of electronic-structure methods and CC3/aug-cc-pVTZ for the oscillator strengths of the 16 carbonyl-containing molecules (each oscillator strength was normalized by the CC3 value). The results for molecules exhibiting a strictly zero oscillator strength for this transition due to symmetry (formaldehyde, acetone, and cyclopropanone) or a very large deviation (cyclobutanone) were omitted from the statistics for clarity.}
     \label{fig:si-normalised}
 \end{figure}

 \begin{figure}[H]
     \centering
     \includegraphics[width=0.7\textwidth]{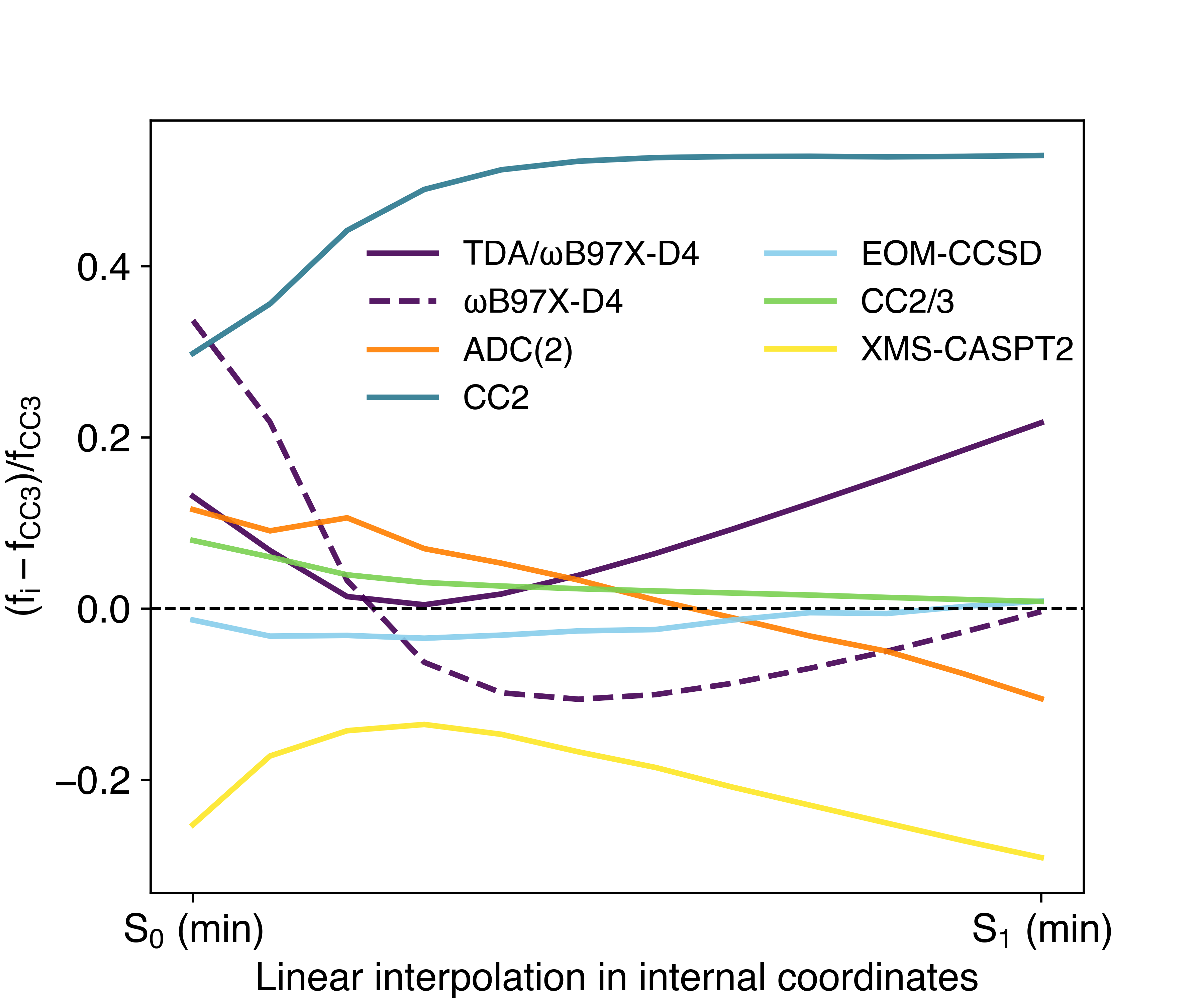}
     \caption{Oscillator strengths between S$_1$ and S$_0$ along the LIIC pathway, represented as a deviation relative to the CC3/aug-cc-pVTZ reference values normalized by the CC3/aug-cc-pVTZ oscillator strength.}
     \label{fig:si-normalised-liic}
 \end{figure}
 
\section{Comparison of various exchange-correlation functionals along the LIIC pathway}

 \begin{figure}[H]
     \centering
     \includegraphics[width=0.95\textwidth]{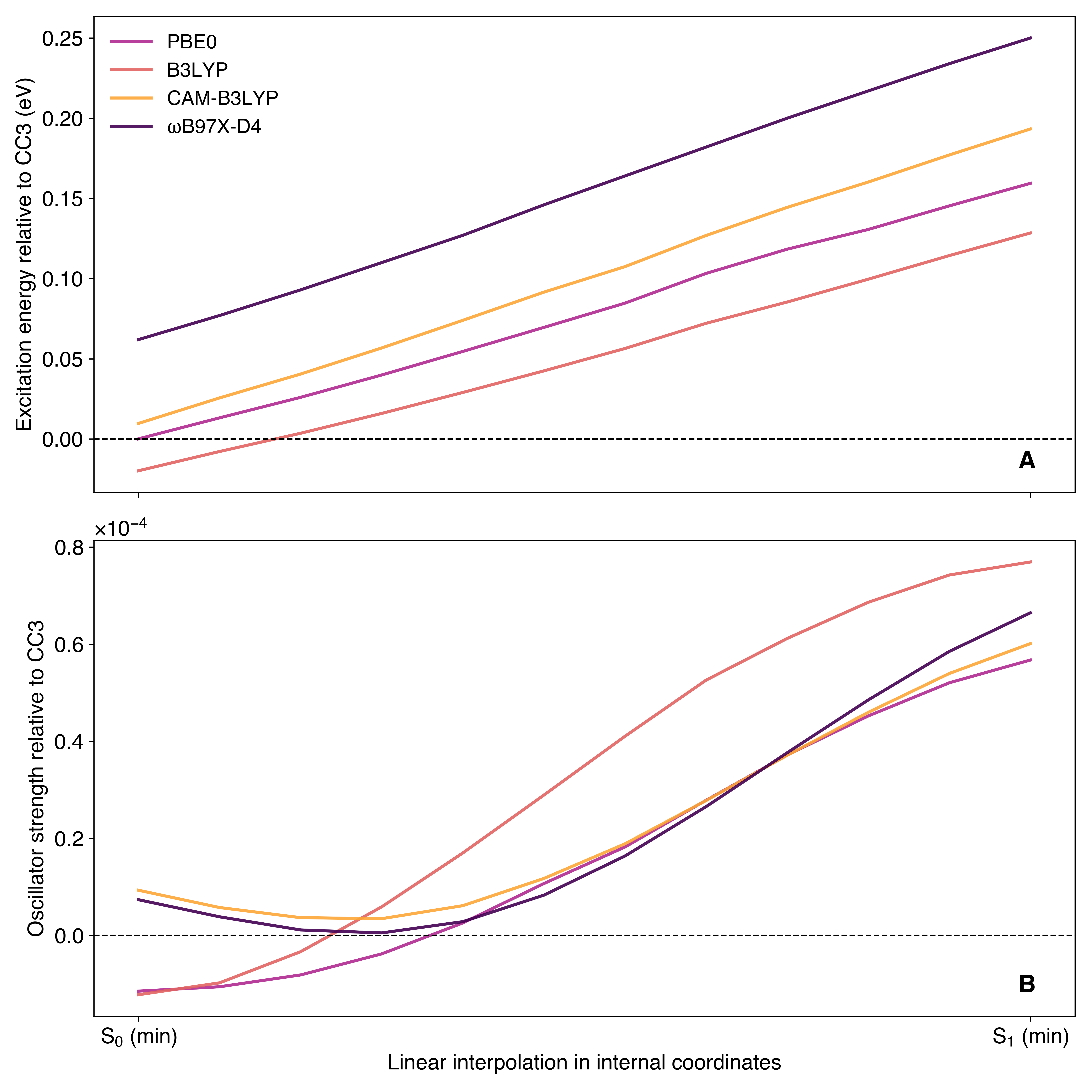}
     \caption{Test of various exchange-correlation functionals with LR-TDDFT/TDA.
    (A) Excitation energies between S$_1$ and S$_0$ along the acetaldehyde LIIC pathway, represented as a deviation relative to the CC3/aug-cc-pVTZ reference values (horizontal dashed line).
    (B) Oscillator strengths between S$_1$ and S$_0$ along the acetaldehyde LIIC pathway, represented as a deviation relative to the CC3/aug-cc-pVTZ reference values (horizontal dashed line). These calculations were conducted with ORCA v6.0.0.}
     \label{fig:si-liic-tda}
 \end{figure}

 \begin{figure}[H]
     \centering
     \includegraphics[width=0.95\textwidth]{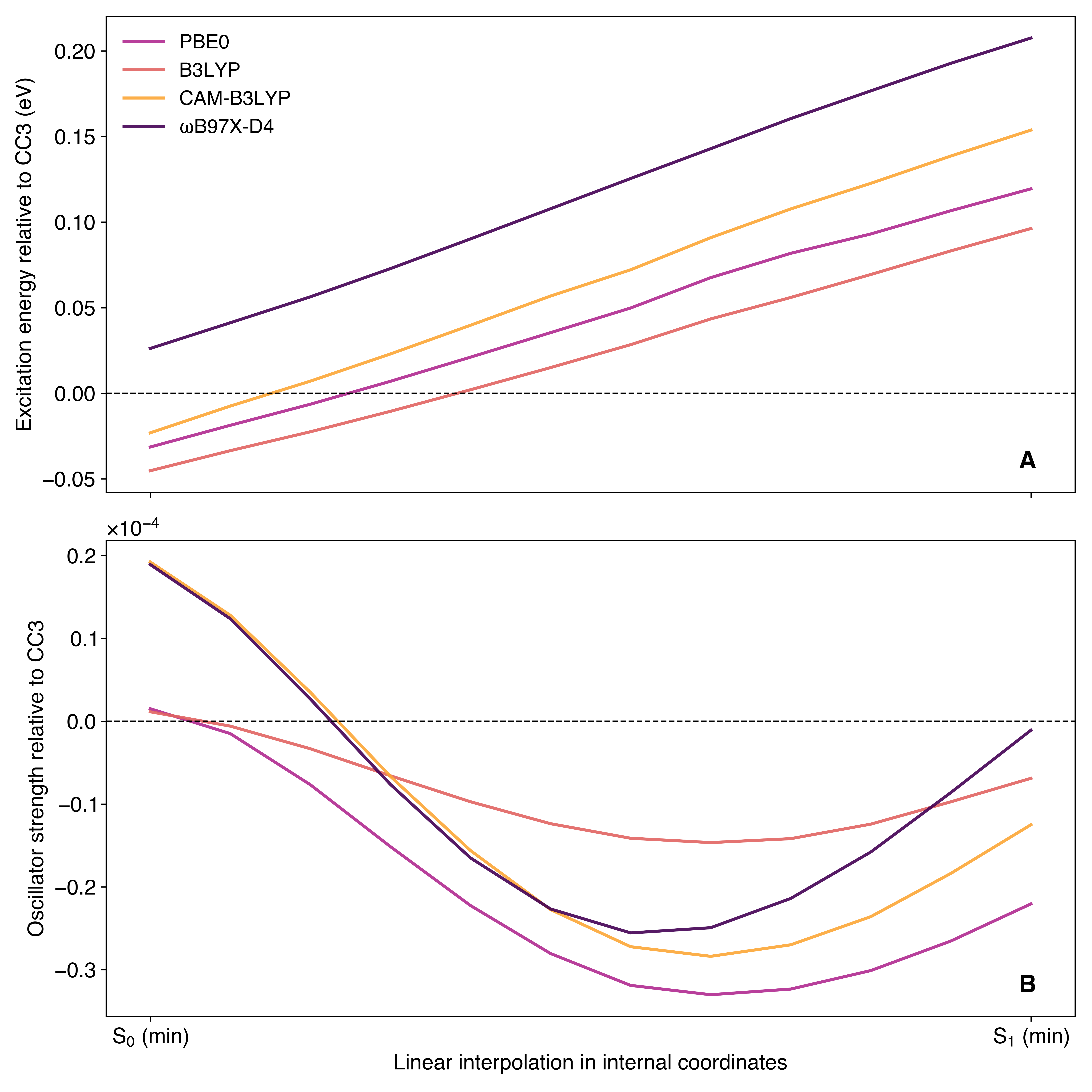}
     \caption{Same as in Figure~\ref{fig:si-liic-tda}, but for LR-TDDFT.}
     \label{fig:si-liic-tddft}
 \end{figure}

\section{Active spaces for XMS-CASPT2 calculations}
\begin{figure}[H]
     \centering
     \includegraphics[width=0.8\textwidth]{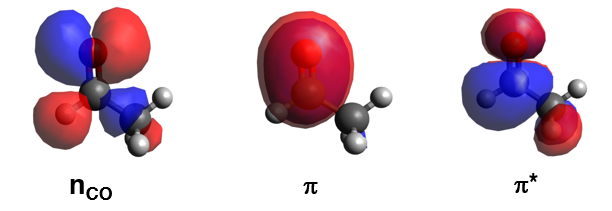}
     \caption{Active space orbitals employed in the XMS(2)-CASPT2(4/3) calculations. Natural orbitals from the SA(2)-CASSCF(4/3)/aug-cc-pVTZ reference wavefunction are given here for acetaldehyde, with an isovalue set to 0.02.}
     \label{fig:si-acetaldehyde-mo}
 \end{figure}

The smallest reasonable choice of active space consisted of 4 electrons in 3 orbitals (\textit{n}, $\pi$ and $\pi^\ast$). This active space was employed for XMS-CASPT2 calculations on acetaldehyde, formaldehyde, acetone, trifluoroacetaldehyde, MEK, cyclobutanone, cyclopropanone, glycolaldehyde, and 3-hydroxypropanal, together with a state-averaging over two electronic states. An example of the natural orbitals constituting this active space can be seen in Figure \ref{fig:si-acetaldehyde-mo} for acetaldehyde. 

\begin{figure}[H]
     \centering
     \includegraphics[width=0.8\textwidth]{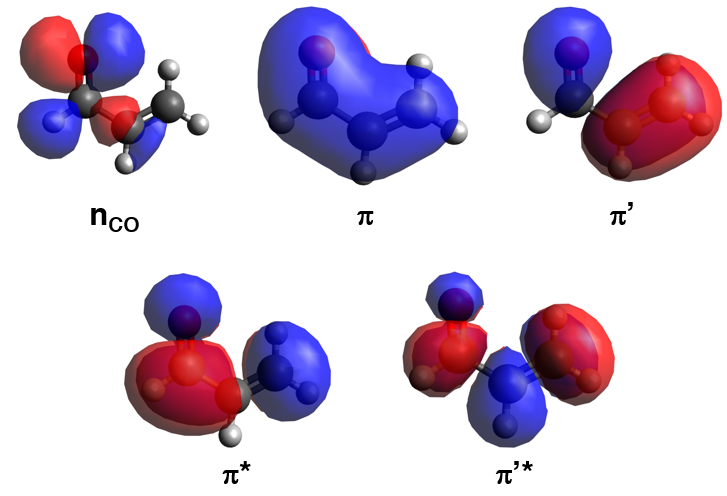}
     \caption{Active space orbitals employed in the XMS(2)-CASPT2(6/5) calculations. Natural orbitals from the SA(2)-CASSCF(6/5)/aug-cc-pVTZ reference wavefunction are given here for acetaldehyde, with an isovalue set to 0.02.}
     \label{fig:si-mvk-mo}
 \end{figure}

Extensions of the (4,3) active space were needed to capture the full excitation character of some molecules in the benchmark set. The (6,5) active space -- used for acrolein and MVK -- includes contributions from the $\pi$/$\pi^\ast$ orbitals of the alkene bond as well as the $n$, $\pi$ and $\pi^\ast$ orbitals of the carbonyl. An example of the natural orbitals in this active space is given in Figure \ref{fig:si-mvk-mo} for MVK(I).

\begin{figure}[H]
     \centering
     \includegraphics[width=0.8\textwidth]{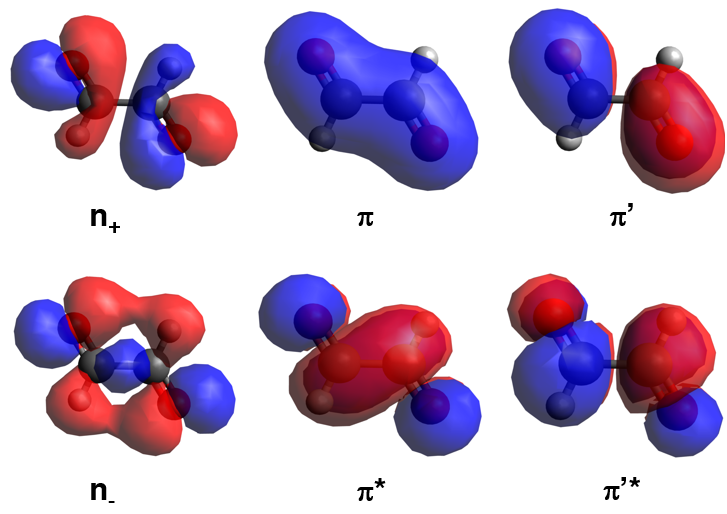}
     \caption{Active space orbitals employed in the XMS(3)-CASPT2(8/6) calculations for glyoxal. Natural orbitals from the SA(3)-CASSCF(8/6)/aug-cc-pVTZ reference wavefunction are plotted here with an isovalue set to 0.02.}
     \label{fig:si-glyoxal-mo}
 \end{figure}

The (8,6) active space was used solely for glyoxal to capture its dialdehyde character. The active space consisted of the $n$, $\pi$ and $\pi^\ast$ orbitals, both in phase and out of phase, to describe adequately the key orbitals for both carbonyls. The natural orbitals included in this active space can be seen in Figure \ref{fig:si-glyoxal-mo}.

For all molecules studied in this work, the $\sigma$/$\sigma^\ast$ orbitals were omitted from the active space. 

\clearpage

\section{Vibronically-resolved photoabsorption cross-section of acetaldehyde}

 \begin{figure}[H]
     \centering
     \includegraphics[width=1.0\textwidth]{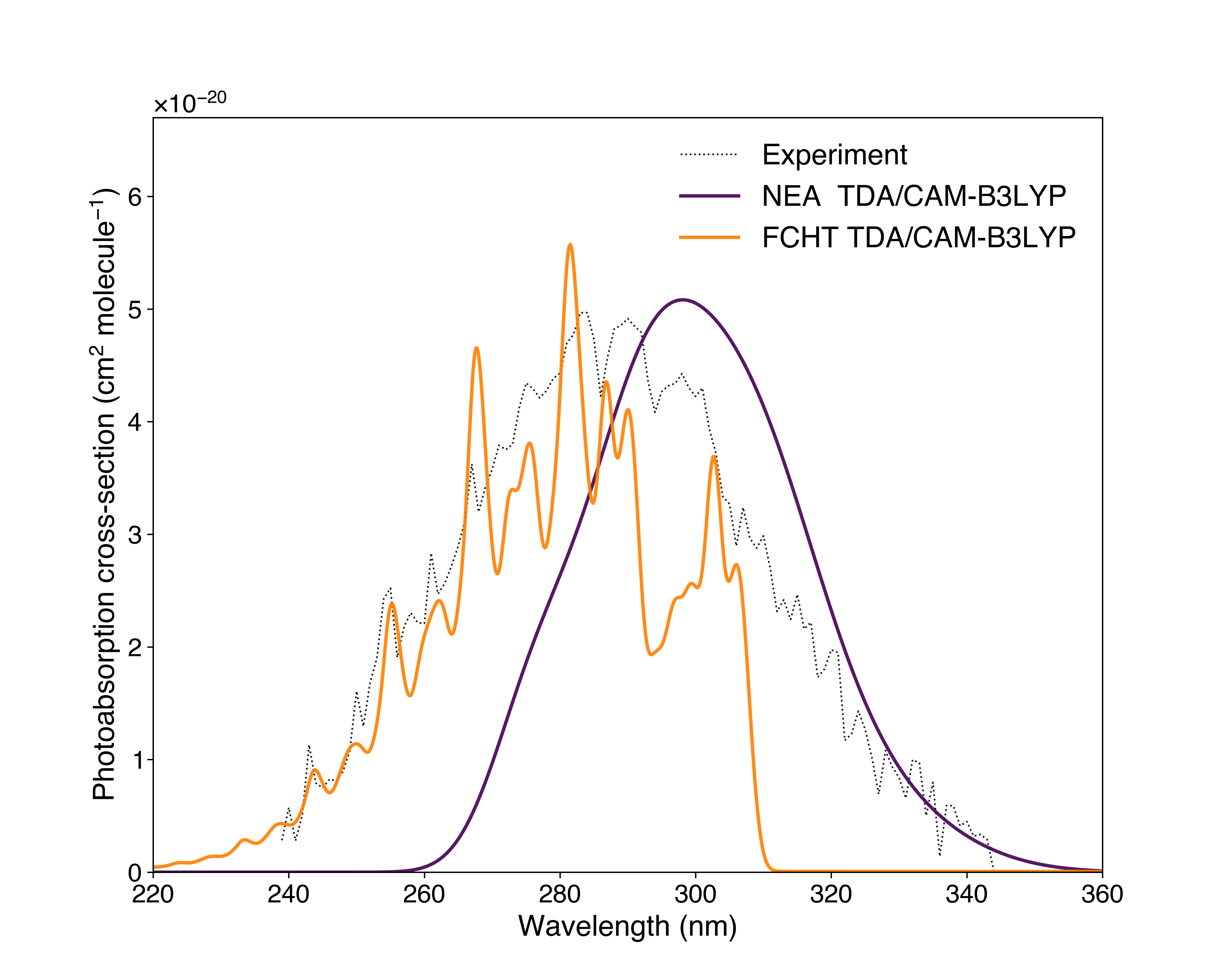}
     \caption{Photoabsorption cross-section of acetaldehyde obtained with the NEA and with the explicit calculation of Franck-Condon Herzberg-Teller (FCHT) factors. LR-TDDFT/TDA/CAM-B3LYP/aug-cc-pVTZ was used for all photoabsorption cross-sections reported. This comparison shows the role played by FCHT factors on the high-energy tail of the photoabsorption cross-section and the limited accuracy of the NEA to reproduce the full photoabsorption cross-section for bound states. The experimental photoabsorption cross-section was presented in Ref.~\citenum{limaovieira2003acetalcross} and obtained via the MPI-Mainz UV/Vis Spectral Atlas.\cite{keller2013mpi}. The FCHT factors were obtained at 0 K using the vertical Hessian (and harmonic) approximation as implemented in ORCA v6.0.0, turning into real any imaginary frequency arising from calculating the Hessian of the final state at a non-stationary point, and convolving each individual transition with a Gaussian function having a half-width at half maximum (HWHM) of 0.02 eV.}
     \label{fig:si-FCHT}
 \end{figure}


\providecommand{\latin}[1]{#1}
\makeatletter
\providecommand{\doi}
  {\begingroup\let\do\@makeother\dospecials
  \catcode`\{=1 \catcode`\}=2 \doi@aux}
\providecommand{\doi@aux}[1]{\endgroup\texttt{#1}}
\makeatother
\providecommand*\mcitethebibliography{\thebibliography}
\csname @ifundefined\endcsname{endmcitethebibliography}  {\let\endmcitethebibliography\endthebibliography}{}